# Well-defined $Cu_2O$ photocatalysts for solar fuels and chemicals

Sourav Rej[1δ], Matteo Bisetto[2δ], Alberto Naldoni[1*] and Paolo Fornasiero[2*]

[1]Regional Centre of Advanced Technologies and Materials, Faculty of Science, Palacký University, Šlechtitelů 27, 78371 Olomouc, Czech Republic.

[2]Department of Chemical and Pharmaceutical Sciences, ICCOM-CNR Trieste Research Unit, INSTM-Trieste, University of Trieste, Via L. Giorgieri 1, 34127 Trieste, Italy.

[δ]Both of these authors contribute equally in this work.

*e-mail: alberto.naldoni@upol.cz ; pfornasiero@units.it

**Abstract**

The shape-controlled synthesis of cuprous oxide ($Cu_2O$) photocatalysts with both low or high index crystal planes has received increasing attention due to their unique facet-dependent properties. Since they are cheap and earth abundant, these well-defined $Cu_2O$ nanostructures are extensively used for different photocatalytic reactions, also because of their strong visible light absorption capability. However, further development will be still needed to enhance the efficiency and photostability of $Cu_2O$, which still limits its industrial application. We start this review by summarizing the synthetic advancement in the facet engineering of $Cu_2O$ and other associated hybrid $Cu_2O$-based heterostructures with a special emphasis put on their growth mechanism. We then discuss different facet-dependent properties, which are relevant to photocatalysis. In the subsequent section, we present a critical discussion on the photocatalytic performance of faceted $Cu_2O$ nanostructures during organic synthesis, hydrogen production, and carbon dioxide photoreduction. The relation between photocatalytic efficiency and product selectivity with exposed crystal facets or with different composition of hybrid nanostructures is also discussed. Finally, important strategies are proposed to overcome the photostability issue, while outlining the course of future development to further boost the technological readiness of well-defined $Cu_2O$-based photocatalysts.

**Contents:**

## 1. Introduction.

Recent development of efficient photocatalytic materials was significantly accelerated in order to effectively deploy green, energy efficient, and sustainable technologies [1–6]. Generally, these materials are inorganic metal oxide semiconductors and, because of their particular band structure, they allow absorption of a broad part of visible light from the solar spectrum. Furthermore, the electronic structure of these materials provides unique optical behavior and conductivity properties [1,3,4]. A typical photocatalytic mechanism consists of three consecutive steps (Fig. 1a): (i) absorption of light with appropriate energy and generation of an electron ($e^-$) hole ($h^+$) pair, which, respectively, (de)localize in the conduction band (CB) and the valence band (VB) of the semiconductor; (ii) electron and hole separation and migration to the catalyst surface in order to participate in the reduction and oxidation reactions, respectively, at the semiconductor-solutions interface; and (iii) if the electron-hole pairs don't take part in the redox reaction, they then undergo the recombination process. The first two steps majorly determine the photocatalytic efficiency of the semiconductor material, whereas the third step is responsible for the performance as such. Thus, to facilitate the solar-to-chemical energy conversion efficiency, different strategies have been developed. They include controlling the morphology, size, and compositions of the photocatalyst for various reactions such as organic reactions, hydrogen production via water splitting (or photoreforming of sacrificial agents), and $CO_2$ reduction [7–14].

The synthesis of inorganic metal oxide nanocrystals with a well-defined structure has been extensively studied over the last decades [15–18]. The opportunity to control the morphology and size of nanocrystals will benefit the light absorbing properties and charge carrier management, thus enhancing the photocatalytic activity and product selectivity in the end [19].

Moreover, different crystal facets possess different atomic arrangements, which may directly control their physical and catalytic properties [20,21]. Therefore, understanding the synthesis and growth mechanism of metal oxide nanocrystals bounded with low or high index facets is really crucial to examining their facet-dependent properties. Common semiconductors like $TiO_2$, ZnO, $WO_3$, and $Cu_2O$ attract significant attention due to their surprising photocatalytic activity caused by their suitable band gap energy (Fig. 1b) [15–17].

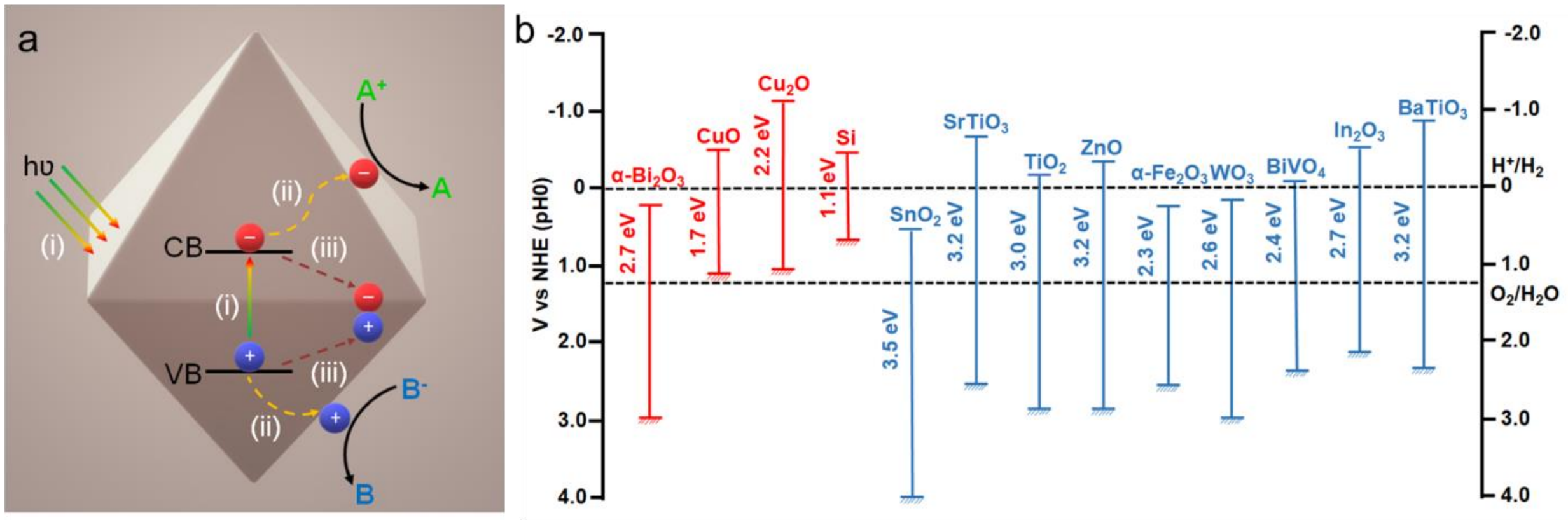

**Figure 1.** (a) Typical mechanism involved in a photocatalytic process. Adapted with permission from ref. [2], Copyright 2018, Wiley-VCH. (b) Band gap values and band energy alignment with water splitting redox potentials of different common *p*-type (red) and *n*-type (blue) semiconductors *vs* NHE and at pH 0. Adapted with permission from ref. [1], Copyright 2013, Wiley-VCH.

Among different photocatalysts, nontoxic, highly abundant cuprous oxide ($Cu_2O$) is a promising photocatalyst for dye degradation, organic reactions, $CO_2$ reduction, and $H_2$ production as it shows a direct band gap structure with a small band gap energy (2.2 eV), which help this material to absorb efficiently in the visible range of the solar spectrum (Fig. 2) [22]. $Cu_2O$ is the most widely synthesised *p*-type semiconductor allowing proper control of its morphology and size obtained under mild synthetic conditions. Therefore, extensive research has been carried out to investigate the $Cu_2O$ facet-dependent photocatalytic properties [23–29]. However, the photocatalytic activity of $Cu_2O$ nanocrystals is still limited by the high recombination rate of the photoexcited electron-hole pairs and also by the photostability issue [30]. The photogenerated holes initiate the self-photooxidation of $Cu_2O$, which causes their decreased photostability [30]. In order to overcome this issue, different hybrid $Cu_2O$-based nanostructures were developed to enhance the photocatalytic activity by concomitantly increasing also the electron-hole separation efficiency. Photoexcited electrons accumulate in the reduction site where they catalyze the reduction reaction; simultaneously, the photogenerated holes are consumed in the oxidation sites by undergoing an oxidation reaction, which increases the photostability of $Cu_2O$ counter parts. Such synergistic behavior in the hybrid system not only enhances the photostability, but it may also increase the product selectivity.

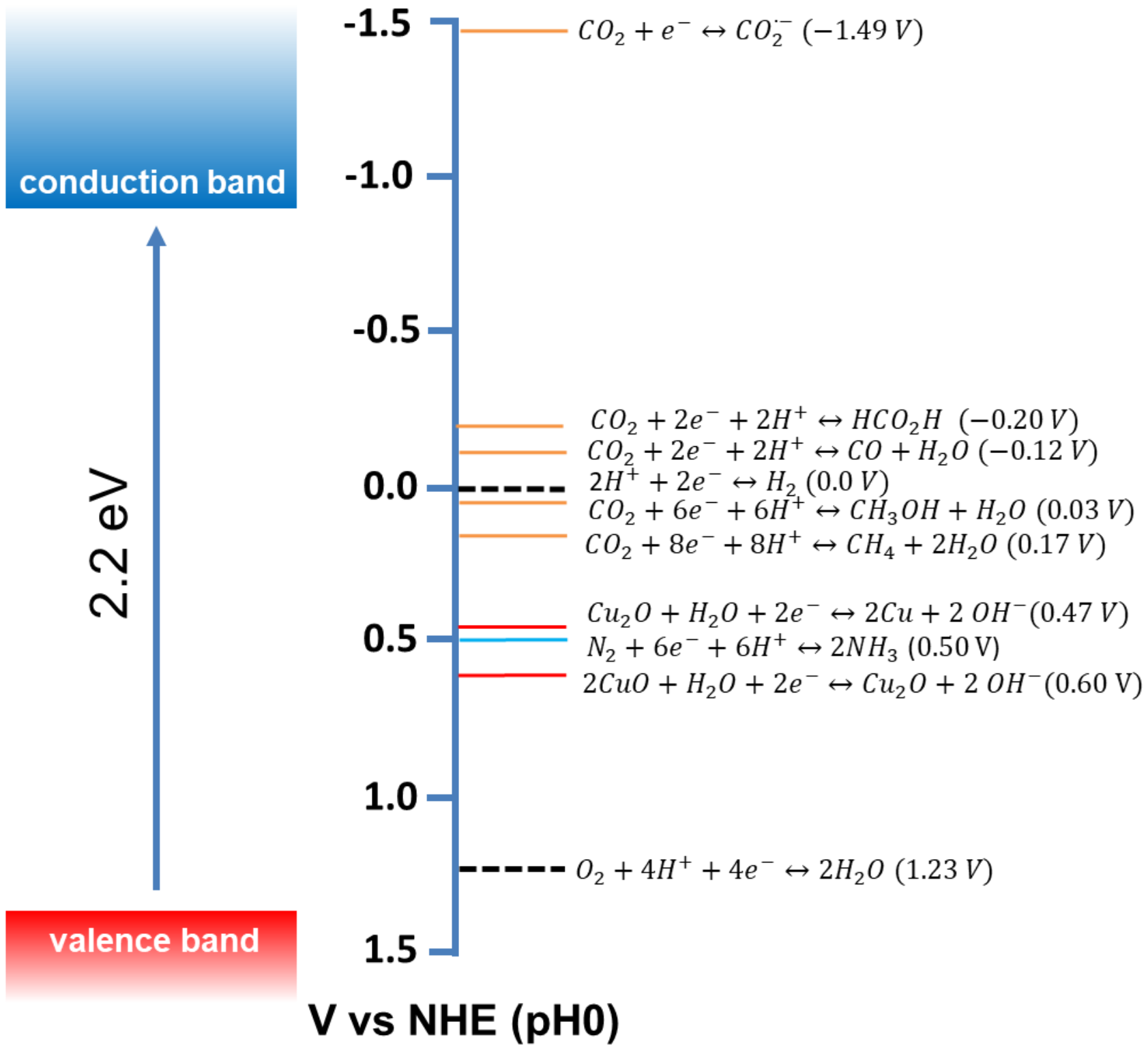

**Figure 2.** Conduction band and valence band potentials of $Cu_2O$ (left side), with the redox potentials of several $CO_2$, $N_2$ and water redox couples at pH 0 plotted *vs* NHE (right side). Adapted with permission from ref. [9], Copyright 2015, American Chemical Society.

Although there are excellent review articles that discuss the synthesis and optical and catalytic properties of well-defined $Cu_2O$ nanocrystals, a comprehensive review focusing on the growth mechanism, surfactant removal process, unique facet-dependent properties, and their photocatalytic application to solar energy conversion is still lacking [23,25–29]. Hence, in this review, we aim to discuss and summarize all the aspects related to the crystal facet engineering of $Cu_2O$ nanocrystals and their photocatalytic application with special emphasis on the stability under different photocatalytic reaction conditions. In the first section, an in-depth discussion on the growth mechanism of well-defined $Cu_2O$ nanocrystal and different synthetic procedures for the preparation of diverse nanostructures is presented. The effect of surfactants and additive ions on controlling the morphology of nanocrystals is also highlighted. Successful strategies for the removal of different kinds of surfactants from nanocrystal surfaces has also been discussed. Next, we present the influence of facets on different chemical properties such as selective absorption behavior, photodeposition of cocatalyst, accumulation of charge carriers on different anisotropic facets, and optical properties. In the last section, we cover the most relevant photocatalytic applications, including $H_2$ production and $CO_2$ reduction, of $Cu_2O$ nanocrystals mentioning various strategies that can be employed to improve the photocatalytic activity and stability of $Cu_2O$ by forming diverse heterostructures.

## 2. Shape & size controlled synthesis of $Cu_2O$ nanocrystals.

This section starts with discussing the growth mechanism for well-defined nanocrystal from crystalline nuclei in a systematic manner. Then, synthetic strategies for diverse $Cu_2O$ nanoarchitectures with precise shape and size control will be presented. The effect of different parameters such as solution pH, concentration of the surfactant, and role of additives will be highlighted. And lastly, tactics for different ways of removing the surfactant without altering the shape and size of $Cu_2O$ nanocrystals will be conferred.

*2.1. General understanding of growth direction and exposed crystal facet.*

Morphology-controlled nanocrystals can be synthesised by different wet chemical synthetic procedures [19]. These chemical reactions often appear to be fairly simple but the exact growth mechanisms inside the glass bottles are extremely complicated. The morphology of the final nanocrystals highly depends on the precise control of temperature, capping agents, different reagents concentrations and also on the sequence of the addition of reagents. Any changes in the aforementioned parameters lead to an unexpected shape and size distribution of the nanocrystals in the reaction mixture. Therefore, scaling up the synthetic processes is not trivial and the industrialization is not easy. The scientific development over the last two decades helped to understand the atomistic details of the evolution pathways in which a precursor compound is converted to atoms, nuclei, and then well-defined nanocrystals. Xia et al. demonstrated that there are three different distinct stages in the growth process: 1) nucleation, 2) evolution of nuclei into seeds, and 3) growth of seeds into nanocrystals [31]. During the first step, in a reductive environment, the concentration of reduced metal atoms steadily increases with time. When the concentration of reduced metal atoms reaches a point of supersaturation, the atoms start to aggregate into small clusters, i.e., nuclei formation takes place. Once formed, these nuclei then grow up to a critical size where further structural modification becomes energetically unfavorable, i.e., the seed formation takes place. These single-crystal seeds usually exist as truncated octahedrons (or Wulff polyhedrons) enclosed by a mixture of {111} and {100} facets as they possess nearly spherical profile, and, thus, the smallest surface area helps to minimize the total interfacial free energy. The density, size distribution, and surface chemistry of these resulting crystalline seeds are highly dependent on the reaction environment of the solution. Once the seeds are formed, the temperature, solvent, reagents, and additives such as surfactant or inorganic ions present in the reaction mixture strongly influence the ultimate growth rate and direction of the seed particles and determine the final morphology of the nanocrystals.

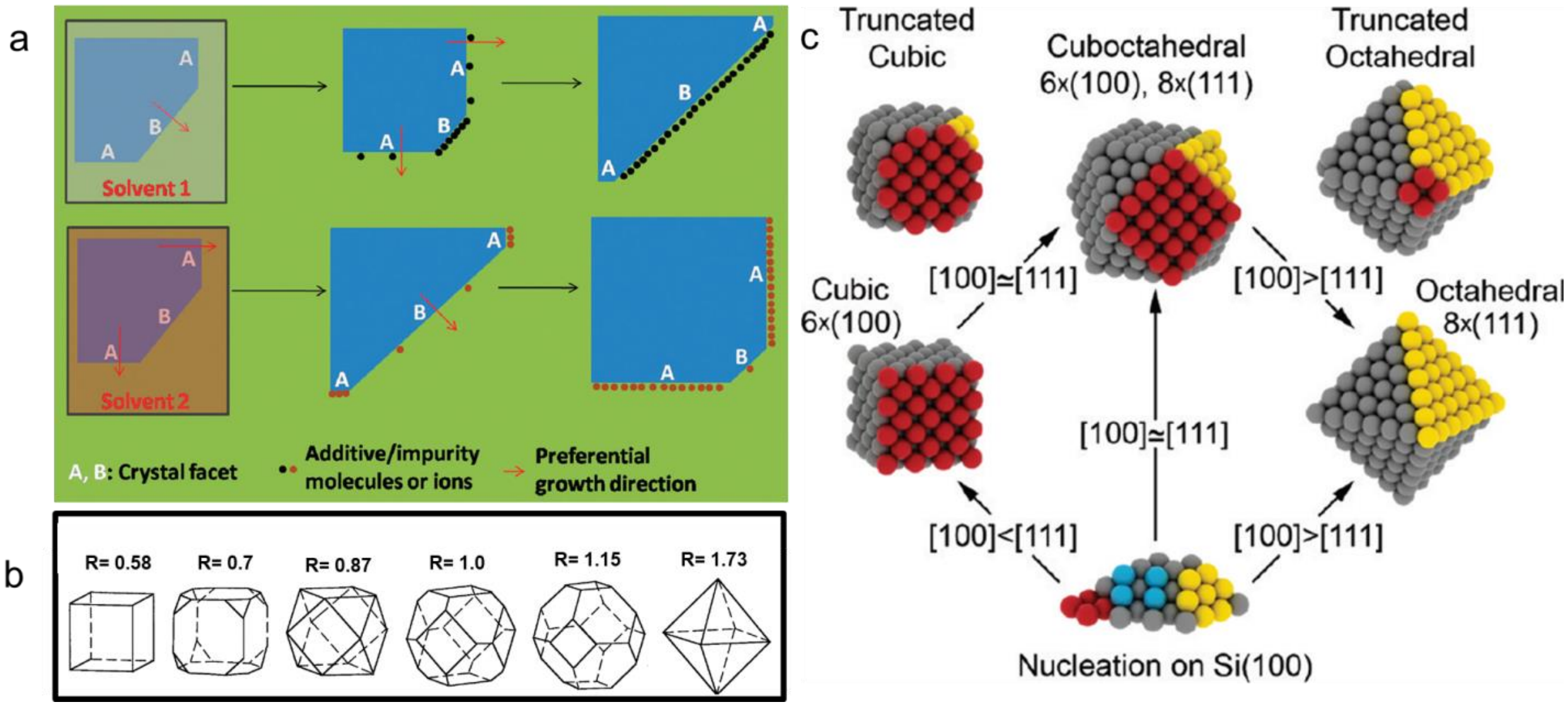

**Figure 3.** (a) Schematic diagram of the growth mechanism of a seed into a well-defined nanocrystal. Reproduced with permission from ref. [26], Copyright 2011, Royal Society of Chemistry. (b) Value of $R_G$ for different morphology. Reproduced with permission from ref. [32], Copyright 2000, American Chemical Society. (c) Scheme for the shape evolution at different $R_G$ values. Red and yellow colored sphere represent the atoms present on (100) and (111) crystal planes in a seed respectively. Reproduced with permission from ref. [33], Copyright 2010, American Chemical Society.

The process of a seed growing to a well-defined nanocrystal follows a route to minimizing the total surface energy which is controlled by both thermodynamics and kinetics of the reaction [31]. Under optimized synthetic conditions, different surfactants, ions or additives present in the reaction mixture selectively adsorb on a specific facet, as it is schematically shown in Fig. 3a.

During the process of growing a seed, the facets which are not blocked by any additives, grow at a higher rate than the unblocked facets [26]. Therefore, the fast-growing facets will eventually disappear, resulting in a nanocrystal terminated with blocked facets. This event tailors the shape of the crystal and enables to expose different facets. According to the Wulff–Kaishew theory, the ratio of the growth rate in the <100> and <111> directions, i.e., $R_G = \frac{r_{<100>}}{r_{<111>}}$ is the deciding factor of the final morphology [32–34]. Based on this, when the value of $R_G = 0.58$ for certain reaction conditions is reached, exclusively cube-shaped nanocrystals are formed and bounded with six low-index (100) crystal planes. This kind of environment favors the fast crystal growth along the <111> direction, compared to <100> direction; therefore, the (111) facets disappear from the seed and the cubic shape comes out as the final product. Similarly, the octahedral shape, bounded with eight low-index (111) crystal planes, becomes dominant when $R_G = 1.73$. Three-dimensional geometric shape evolution from cubic to octahedral morphology with different $R_G$ values is shown in Fig. 3b. When $0.58 < R_G < 1.73$, different truncated forms of nanocrystals can be obtained in the reaction medium (Fig. 3c). Therefore, it can be concluded that the final shape of the nanocrystal and the exposed facets are the result of the interplay between thermodynamics and kinetics of the reaction.

Although there is enough theory and experimental routes to synthesizing well-defined nanocrystals, the atomic pathways of nanocrystal facet development have been mostly unknown to date because of the lack of direct microscopic observations inside the reaction mixture. Currently, in-situ liquid cell transmission electron microscopy (LCTEM) allows observing the single-nanoparticle growth trajectories. A liquid cell contains a small amount of liquid reaction mixture inside the high-vacuum microscope and the nanocrystal growth can be initiated by the introduction of an electron beam that often acts as a reductant. Wu et al. observed the growth process of a $Cu_2O$ nanocube by reducing an alkaline copper precursor solution with an electron beam [35]. The time-dependent growth process of a $Cu_2O$ nanocube observed via in-situ LCTEM is shown in Fig. 4. When the electron beam is introduced into the precursor solution, the $Cu^{+2}$ ions become reduced to $Cu^{+1}$ ions and initiate the $Cu_2O$ nucleation in the presence of hydroxide anion, as shown by the darker region in Fig. 4a, which then eventually is converted to a nanocube via a complex mechanism. Further development will be needed in this field to gain knowledge of the morphology-controlled nanocrystal synthesis [36].

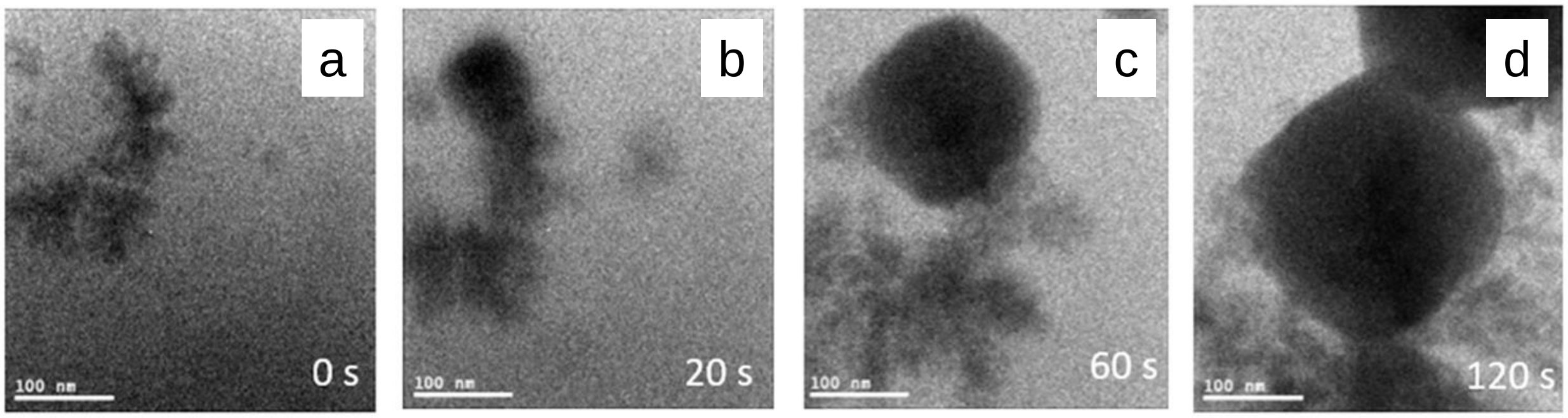

**Figure 4.** Time-dependent in-situ TEM images to track the growth of a single $Cu_2O$ nanocube in a reaction mixture. Reproduced with permission from ref. [35], Copyright 2019, American Chemical Society.

*2.2. Synthesis of diverse $Cu_2O$ nanostructures.*

A large number of synthetic wet-chemical routes to producing a wide variety of well-defined $Cu_2O$ nanocrystals have been developed over the last two decades. To better understand the effect of the role of different reagents in controlling the morphology, different relevant case studies are presented in this section with proper explanation.

Murphy et al. developed a synthetic procedure of $Cu_2O$ nanocubes using an aqueous solution of copper sulfate ($CuSO_4$), NaOH, and cetyltrimethylammonium bromide (CTAB) as the surfactant and sodium ascorbate as the reducing agent [37]. The shape and size of the nanocrystals can be tuned by varying the concentration of the surfactant. At low surfactant concentrations, the nanocrystals are ineffectively capped and grow randomly with a spherical morphology. When an optimum concentration is reached, it leads to adequate surface capping, and uniform cubic nanoparticles are formed. TEM images confirm that these crystals appear to be hollow from inside, which is the major drawback for this synthetic condition. At that time, Wang group also developed a procedure to synthesize $Cu_2O$ nanocubes and octahedral shape [38]. Typically, an $NH_3$ solution is first added to an aqueous $CuCl_2$ solution followed by adding NaOH, which leads to the blue precipitation of $Cu(OH)_2$. After that, an aqueous solution of $N_2H_4$ is added as the reducing agent to the above-described solution to start the formation of $Cu_2O$ nanocrystals. The molar ratios of reagents, i.e., $R_1 = \frac{[NH_3]}{[Cu^{+2}]}$ and $R_2 = \frac{[OH^-]}{[Cu^{+2}]}$ determine the morphology of the corresponding products by affecting the coordination between $NH_3$ and $Cu^{+2}$ ions, which makes this report interesting. When $R_1$ = 0, $R_2$ = 2, i.e., there is no ammonia present in the system, porous $Cu_2O$ spheres were obtained as this morphology has the lowest surface energy. Keeping $R_2$ = 2 fixed, cube-like and octahedral $Cu_2O$ nanocrystals were formed when $R_1$ = 4 and 7, respectively. This proved that ammonia plays a critical role in this reaction in order to control the final morphology of the $Cu_2O$ nanocrystals. In the beginning, $Cu^{+2}$ ions undergo complexation with $NH_3$ molecules forming $[Cu(NH_3)_4]^{+2}$; when NaOH is added, $NH_3$ molecules are replaced by $OH^-$ ions to form a blue $Cu(OH)_2$ precipitate. Under fixed $R_2$ value, the maximum number of $NH_3$ molecules are replaced by $OH^-$ ions when $R_1$ = 4, whereas the minimum number of $NH_3$ molecules are replaced when $R_1$ = 7. This phenomenon leads to distinct pH values under the two different synthetic conditions. Therefore, $R_G$ values (i.e., the ratio of growth rates along the <100> and <111> direction) become 0.58 and 1.73 when $R_1$ = 4 and $R_2$ = 7, resulting in cube-like and octahedral nanocrystals, respectively, in the reaction medium. These synthetic conditions provide a stimulating idea on how we can control the shape of the nanocrystals by the fine pH tuning of an inorganic reaction without adding any organic surfactants. However, this approach mainly suffers from inhomogeneity in the size and shape of the nanocrystals, and, therefore, it requires further development.

To address this problem, Huang et al. developed a strategy to synthesize $Cu_2O$ nanocrystals with systematic shape evolution by a single chemical reduction method at room temperature [39]. Cubic, truncated cubic, cuboctahedral, truncated octahedral, octahedral, short and extended hexapod structures of $Cu_2O$ nanocrystals with systematic morphological evolution have been synthesized by sequential addition of the sodium dodecyl sulfate (SDS) surfactant, an aqueous solution of $CuCl_2$, NaOH, and hydroxylamine hydrochloride ($NH_2OH{\cdot}HCl$) as reductants. The mixture aged for 2 h to obtain the products. The systematic variation in the product morphology can be obtained by changing the amount of added $NH_2OH{\cdot}HCl$, as shown in Fig. 5a–h. Different pH values obtained for different morphologies suggest clear correlation between the pH and the $R_G$, which proportionally controls the final shape of the nanocrystals, as shown in Fig. 5. Keeping other parameters fixed, increasing the amount of $NH_2OH{\cdot}HCl$ can make the solution pH decrease from 12.00 for the nanocube sample to 8.70 for the short hexapod sample. Apart from SEM images, the high morphological uniformity of these $Cu_2O$ nanocrystals can be associated with their PXRD patterns, as shown in Fig. 5j. The PXRD patterns evidently show a change in the relative intensities of the (111) and the (200) peaks, which also goes along with a with morphology change. Nanocubes show only a strong (200) reflection peak with an extremely weak (111) reflection peak. The intensity of the (111) peak increases progressively as nanocrystals with more {111} surfaces are formed. The two peaks become more comparable in intensity for the type I truncated octahedra. The (111) peak then dominates for octahedra and hexapods.

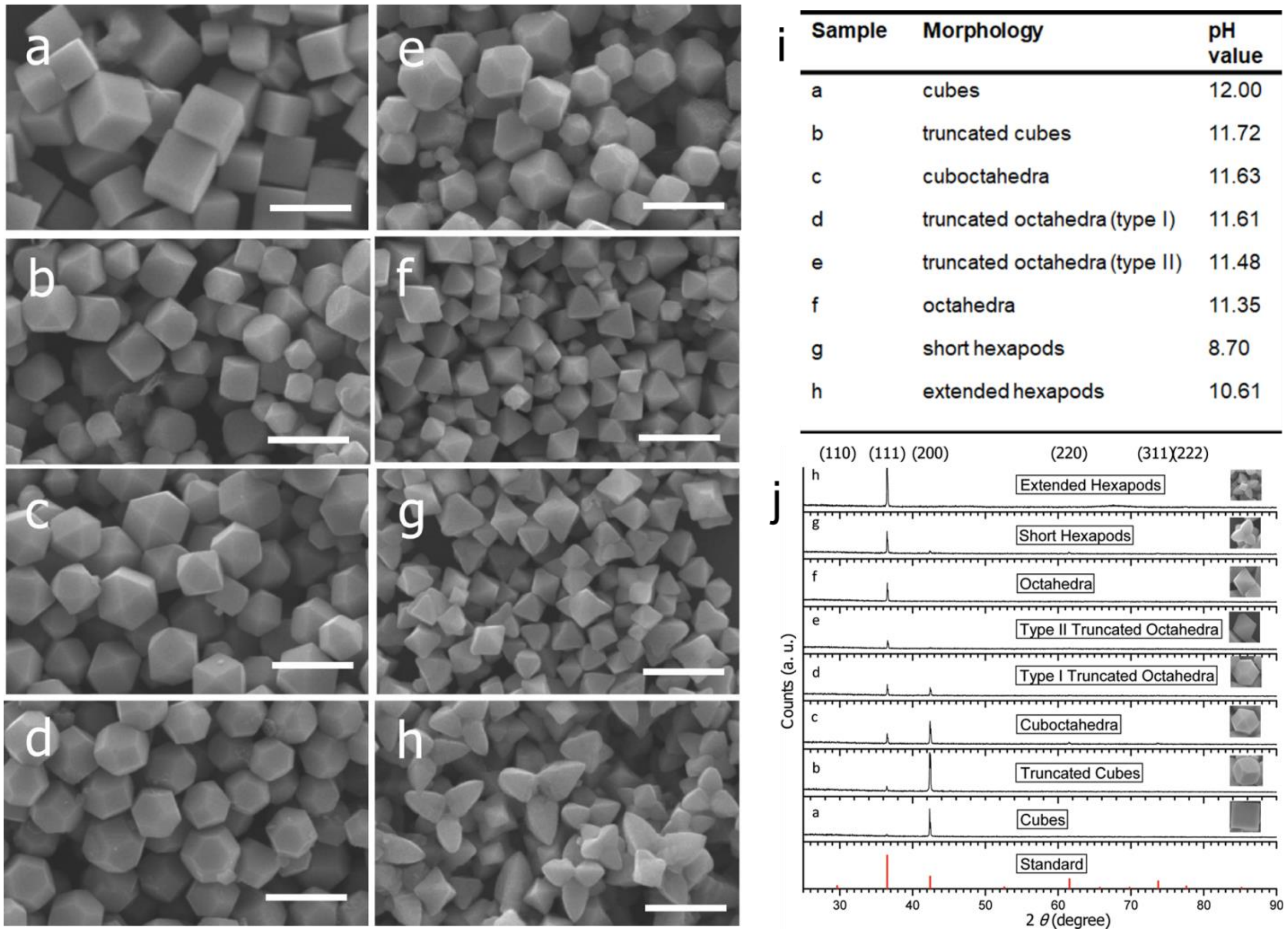

**Figure 5.** SEM images of the $Cu_2O$ nanocrystals with various morphologies: (a) cubes, (b) truncated cubes, (c) cuboctahedra, (d) type I truncated octahedra, (e) type II truncated octahedra, (f) octahedra, (g) short hexapods, and (h) extended hexapods. Scale bar = 1 μm. (i) Solution pH values for samples a–h. (j) PXRD patterns of the different shaped $Cu_2O$ nanocrystals as shown in a-h. Reproduced with permission from ref. [39], Copyright 2009, American Chemical Society.

Apart from controlling the pH of the solution, the surfactant concentration can also play a crucial role in controlling the systematic shape evolution of the nanocrystals. Polyvinylpyrrolidone [PVP] is one of those surfactants that not only act as stabilizers to prevent the aggregation of nanocrystals but also assists the formation of well-defined $Cu_2O$ nanocrystals [40]. PVP molecules with long chains can be adsorbed on the $Cu_2O$ surfaces via both physical and chemical bonding. In particular, PVP preferentially interacts more strongly with the {111} facets than the {100} facets, which suppress the growth of the {111} planes more efficiently than the {100} planes. Thus, as shown in Fig. 6, when the PVP concentration is low (0.5 mM) or the PVP is absent in the reaction mixture, the capping effect of PVP toward $Cu_2O$ nanocrystals is the weakest or absent, which leads to the formation of cubic shapes. By contrast, when the PVP concentration is high (4.5 mM), it strongly coordinates with the {111} facets and efficiently lowers the surface energy of the corresponding crystal planes. This enhanced coordinating effect of PVP completely blocks the growth on the {111} facets and facilitate the growth on the {100} facets, as shown previously in Fig. 3a. As the growth process takes place rapidly on the {100} facets, they disappear completely, and perfect octahedral morphology appears. Furthermore, different truncated octahedra as intermediate products can be achieved through precise control of PVP concentrations between the 0.5–4.5 mM range. Interestingly, when the PVP concentration becomes 9 mM, spherical nanocrystals are formed due to the high coverage of PVP on all the planes of $Cu_2O$ nanocrystals as there is no preferential growth of any particular crystals plane over another.

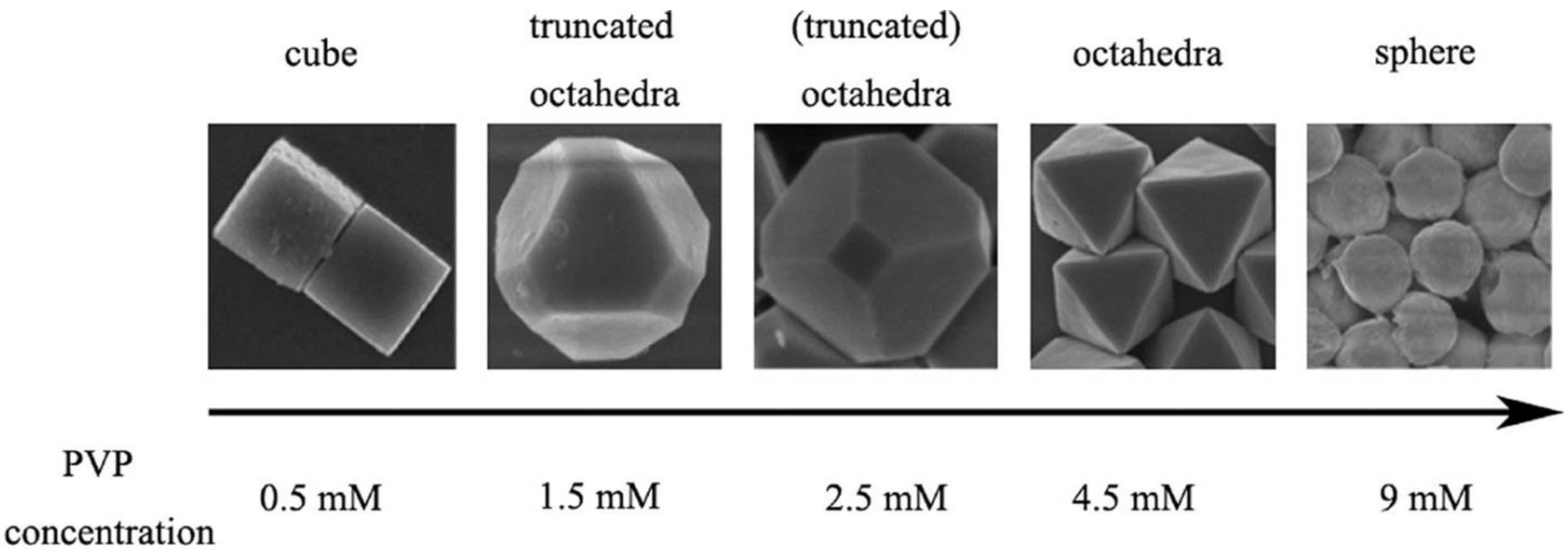

**Figure 6:** Effect of PVP concentration on the $Cu_2O$ morphology. Reproduced with permission from ref. [40], Copyright 2010, American Chemical Society.

Similar to PVP, oleic acid also shows facet-selective adsorption properties, which provides the control over the final morphology of $Cu_2O$ microcrystals [41]. Guo et al. demonstrated the reduction of $CuSO_4$ at 100 °C for 1 h with D-(+)-glucose in an aqueous NaOH/ethanol/oleic acid system to synthesize $Cu_2O$ microcrystals. In the presence of 1mL, 2.5 ml, and 4 ml of oleic acid, uniform cubic, octahedron, and rhombic dodecahedral (RD), $Cu_2O$ microcrystals were obtained, respectively. This was the first report showing the synthesis of $Cu_2O$ RD microcrystals fully bounded by {110} planes, which had a higher surface energy than the corresponding {100} and {111} planes. Later on, Zeng et al. obtained highly uniform ~70 nm $Cu_2O$ RD nanocrystals by the reaction of copper(II) acetate, hexadecylamine, and undecane at 200 °C for 90 min [42]. Hexadecylamine played multiple roles in this study, serving as: 1) a chelating ligand to form $[Cu(NH_2C_{16}H_{33})_4]^{+2}$ complex precursor with $Cu^{+2}$ ions, 2) a phase-transferring agent to transfer divalent $Cu^{+2}$ ions into the organic phase, 3) a reducing agent to generate $Cu^{+1}$ ions, 4) a surface passivating adsorbate to control the crystal morphology. In addition, ~9 μm $Cu_2O$ RD microcrystals were also synthesized by mixing $Cu(NO_3)_2$, formic acid, and $NH_3$ in an ethanol-water solution followed by hydrothermal treatment at 145 °C for 90 min [43].

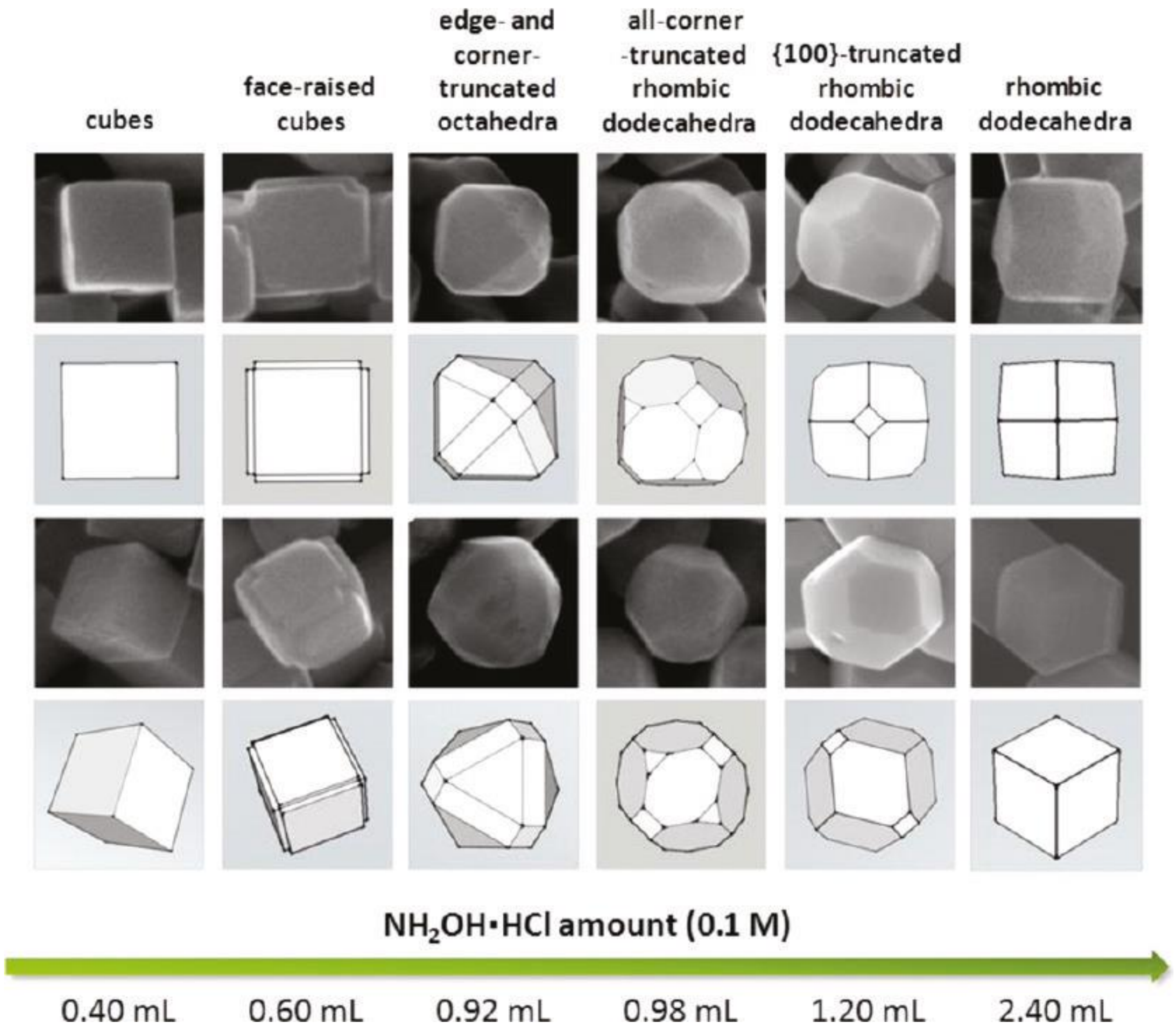

**Figure 7:** Shape evolution from cubic to rhombic dodecahedral morphology with respect to the added amount of hydroxylamine hydrochloride ($NH_2OH.HCl$). Reproduced with permission from ref. [44], Copyright 2012, American Chemical Society.

The major drawback of these reaction conditions is that they need high temperature and strong surface binding ligands, and, therefore, further development will be needed to synthesize the RD shape at room temperature using an easily removable surfactant of a considerably small size. Huang et al. then proposed a synthetic condition that would overcome all the above-mentioned problems [44]. A systematic shape evolution from cubic to RD nanocrystals was obtained at room temperature by mixing an aqueous solution of $CuCl_2$, SDS surfactant, NaOH, and $NH_2OH.HCl$ as the reductant. The systematic shape evolution was achieved only by adjusting the volume of the reducing agent, as shown in Fig 7. The shape evolution is directly related to the final solution pH and overall reduction rate of the copper precursors. It was demonstrated that cubes are formed at a much faster rate than rhombic dodecahedra, thus linking the formation of different particle morphologies to their different growth rates. This protocol is highly important as it allows achieving a diverse nanocrystal morphology with anisotropic facets.

In addition to the role of the concentration of the surfactant and the reducing agent, different inorganic ions are also crucial to controlling the morphology of the nanocrystal. Xia et al. demonstrated a high temperature ethylene glycol (EG) reduction method for synthesizing $Cu_2O$ cubic nanostructures in the presence of chloride ions as the shape directing agent (Fig. 8) [45]. The in-situ formed CuCl intermediate serves as a reservoir to control the supersaturation concentration of the $Cu^{+1}$ ions in the reaction mixture. Therefore, the $Cu_2O$ formation rate significantly slowed down, allowing the seeds to grow into small individual nanocrystals without a significant aggregation at the early stages of the reaction. Chloride ions also stabilize the {100} planes of the $Cu_2O$ nanocrystals and favors the formation of single-crystalline nanocubes as seeds, which can further grow in size via the Ostwald ripening process. When no chloride ions were added to the reaction mixture, the nucleation and growth of $Cu_2O$ nanocrystals took place instantaneously. This uncontrolled growth rate results in the formation of nanoparticles of small size that agglomerated to become polycrystalline colloidal spheres of large sizes in order to reduce the surface energy. Similarly, by tuning the ratio of the concentration of hydroxide and citrate anions, cubes to complete the octahedral-morphological evolution for $Cu_2O$ nanocrystals were obtained on conductive substrates by a method of simple chemical deposition [34]. Feng et al. also used the cyclic scanning electrodeposition (CSE) method to convert a thin Cu film into different $Cu_2O$ structures with the help of different ions such as $NO_2^-$, $NO_3^-$, $SO_4^{-2}$, and $CH_3COO^-$ used in the electrolyte [46]. Other noticeable examples in this regards can also be seen in the literature [47–53].

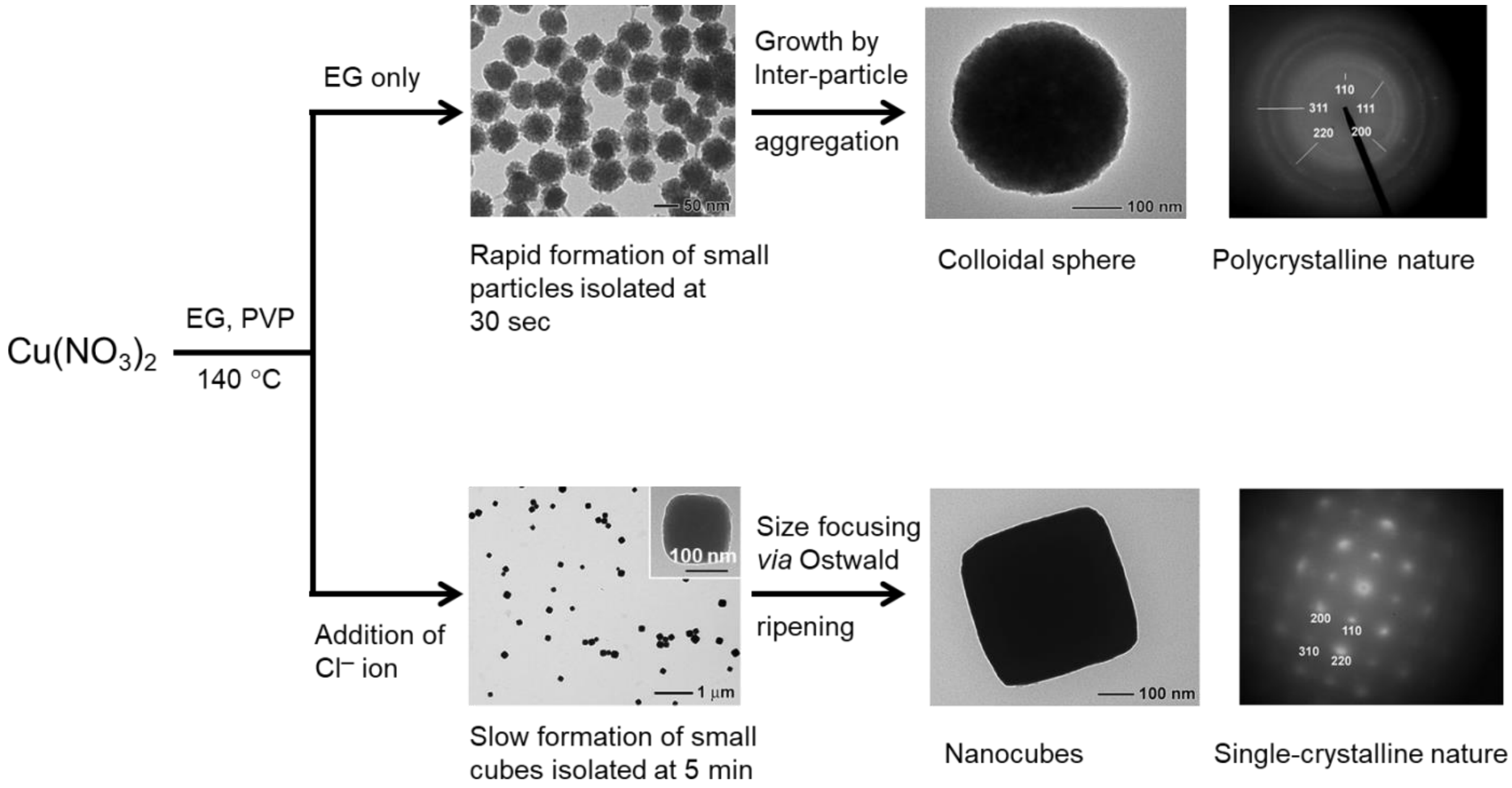

**Figure 8.** Effect of $Cl^-$ ions in controlling the morphology of $Cu_2O$ nanocrystals. Reproduced with permission from ref. [45], Copyright 2008, Royal Society of Chemistry.

After an in-depth discussion on the development of $Cu_2O$ polyhedral structures containing three different low-index facet such as {100}, {111} and {110}, it is necessary to understand the formation mechanism of high-index {hkl} facets (where at least one of the h, k, and l is equal to two or above). Surface energies ($\gamma$) corresponding to different $Cu_2O$ crystallographic facets usually increase in the order of $\gamma$ {100} < $\gamma$ {111} < $\gamma$ {110} < $\gamma$ {hkl}. As these high-index facets possess higher surface energy than the corresponding low-index facet, they disappear faster during the crystal growth and are difficult to preserve on the surface of the final nanocrystal [16]. The unique feature of these facets is that they consist of high density of low-coordinated atoms such as steps, edges, and kinks, which show higher catalytic activity by quickly coordinating with reagent molecules when compared to the low-index facets [16]. Therefore, it is useful to understand the synthetic recipes and the growth mechanism for the preparation of $Cu_2O$ nanocrystals with high-index facets [54–60].

Yang et al. synthesised highly symmetric multi-facet polyhedral $Cu_2O$ microcrystals partially enclosed with high-index facets (i.e., including low-index {110}, {100}, and {111} facets and high-index {544}, {522}, and {211} facets) in the system of $Cu^{+2}$/NaOH/glucose or ascorbic acid solution at high temperature (Fig. 9a1−c2) [54]. There are also other methods of synthesizing $Cu_2O$ microcrystals with high index facets [55–59]. The major drawback of the aforementioned procedures is that the size of $Cu_2O$ crystals is on the order of several microns, which means a very low specific surface areas and eventually lower photocatalytic activity. To solve this problem, Li et al. synthesized $Cu_2O$ truncated concave octahedral nanocrystals of an average edge length of 158 nm, which enclosed {511} high-index facets together with {110} and {100} facets through a system of oil in a water emulsion, as shown in Fig. 9d1−d4 [60]. Both the solution pH and the oleic acid are critical for the formation of truncated concave octahedral $Cu_2O$ nanocrystals.

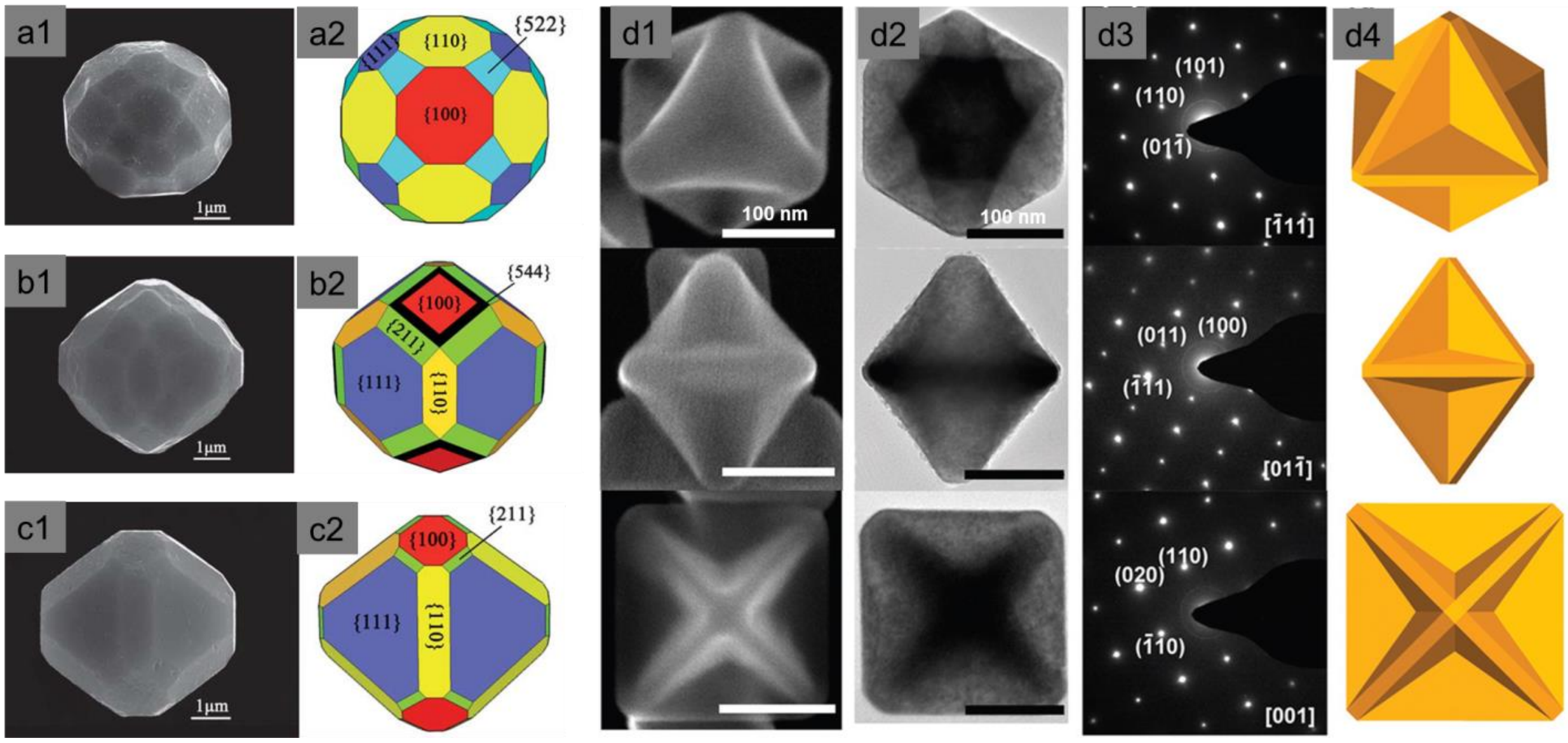

**Figure 9:** Typical FESEM images of polyhedral $Cu_2O$ microcrystals (a1–c1) 50-facet, 74-facet, and 50-facet with high-index {522}, {544} and {211} facets respectively; (a2–c2) representative graphical structure. Reproduced with permission from ref. [54], Copyright 2008, Royal Society of Chemistry. Truncated concave octahedron nanocrystals and its (d1) SEM images, (d2) TEM images, (d3) selective area electron diffraction (SAED) patterns, and (d4) represented models. Reproduced with permission from ref. [60], Copyright 2008, Royal Society of Chemistry.

All the above-mentioned well-defined $Cu_2O$ nanocrystals with both low- and high-index facets have one common drawback, i.e., their size is larger than 100 nm or 1000 nm, depending upon different synthetic conditions. That means they possess a low surface area for the given weight of the catalyst, which is related to a lower catalytic activity. Therefore, it is a challenging task to develop a scalable synthetic method for ultrasmall and highly uniform $Cu_2O$ polyhedral, keeping the facet intact. Several early trials were made to achieve this goal [61–64]. By the reaction of an aqueous mixture of $CuSO_4$, trisodium citrate, and NaOH,  40 nm $Cu_2O$ nanocubes were synthesized, but the size of the particles was still not sufficiently small and therefore the synthetic procedure provided low yields [61]. Huang's group have recently developed a simple method to produce $Cu_2O$ nanocubes and octahedra with a varying size from 16–86 nm and 34–49 nm, respectively (Fig. 10) [64]. The reaction was carried out at room temperature and lasted 20 min. In a typical reaction, $CuSO_4$, NaOH, and sodium ascorbate are added stepwise to an aqueous SDS surfactant solution. A gradual increase in the amount of sodium ascorbate helped to achieve $Cu_2O$ nanocubes of smaller sizes; this is due to the formation of higher number of uniform nucleation centers at higher concentrations of the reducing agent. There are still a few challenges that need to be overcome in order to achieve RD and other nanostructures with a high-index facet within the sub-40 nm range, which can enhance the catalytic activity by increasing the reactive surface area. Further, the sub-40 nm nanocrystals will have more light absorption capabilities than scattering, which will increases their photocatalytic performances.

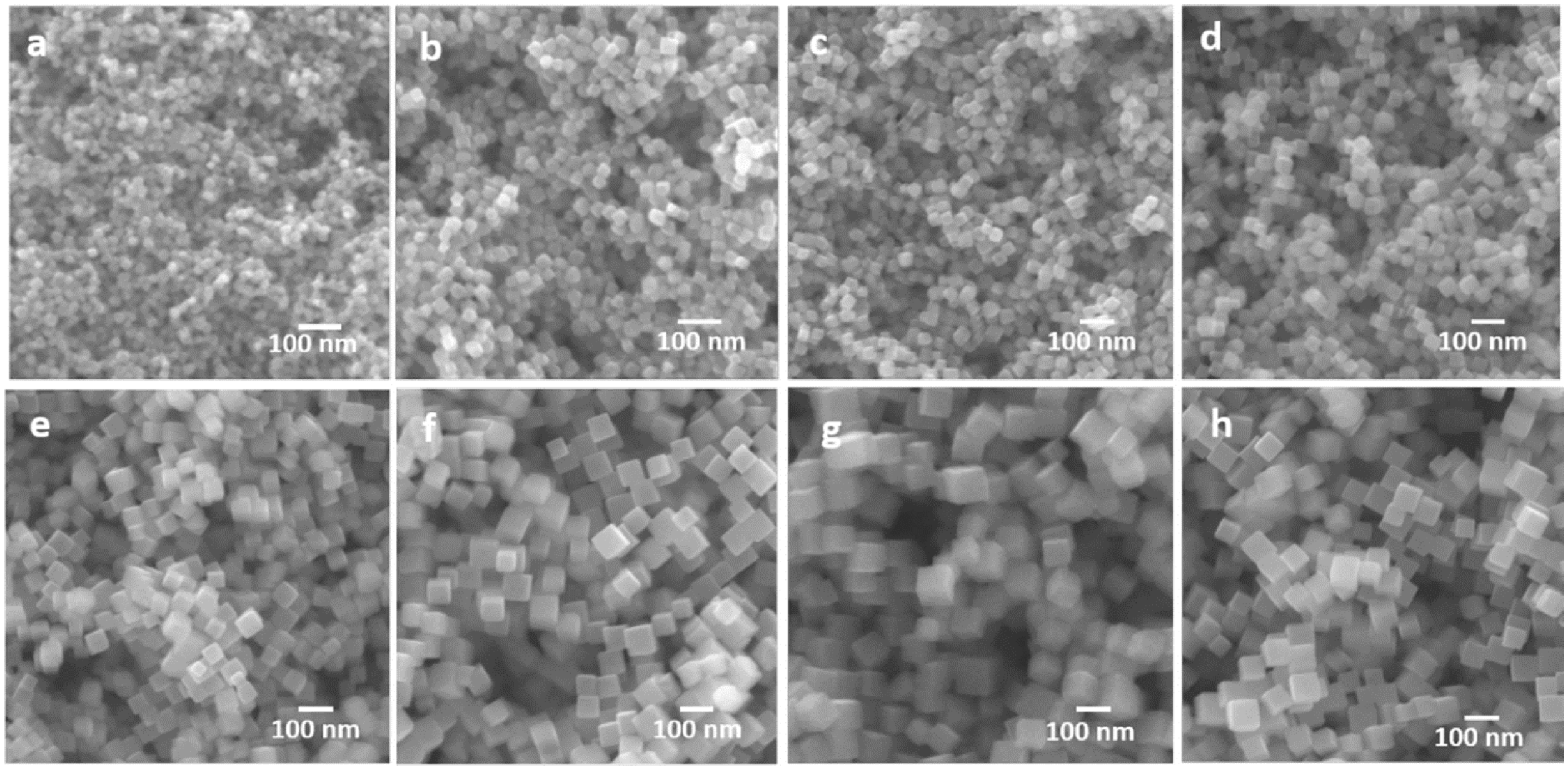

**Figure 10:** Size controlled synthesis of uniform $Cu_2O$ nanocubes with average edge lengths of (a) 16, (b) 25, (c) 29, (d) 36, (e) 51, (f) 63, (g) 72, and (f) 86 nm. Reproduced with permission from ref. [64], Copyright 2019, American Chemical Society.

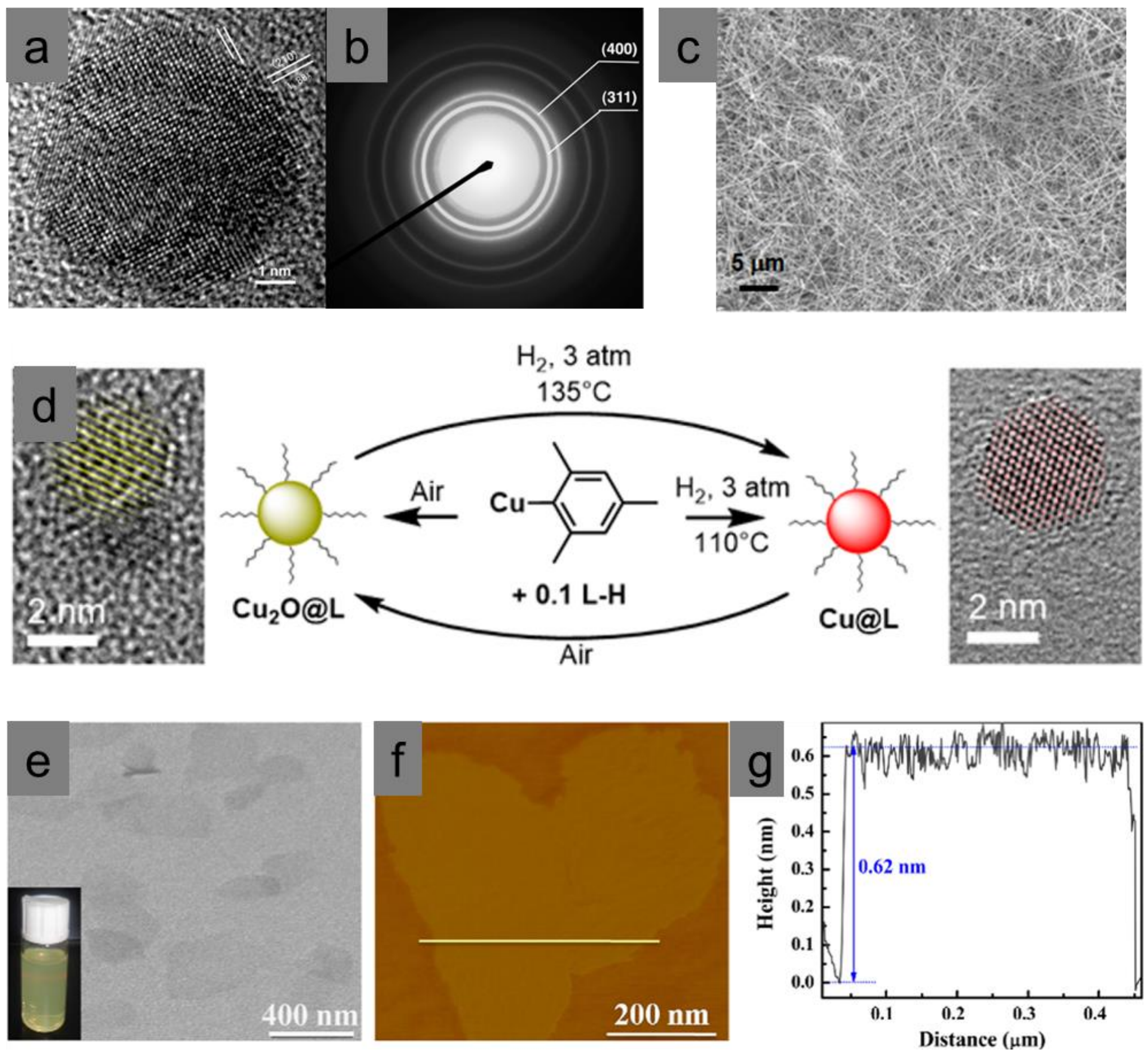

**Figure 11:** Highly single crystalline $Cu_2O$ nanocrystals (a) HRTEM image and its (b) SAED pattern. Reproduced with permission from ref. [65], Copyright 2005, American Chemical Society. (c) SEM image of $Cu_2O$ nanowires. Reproduced with permission from ref. [66], Copyright 2007, American Chemical Society. (d) Reversible synthesis of ultrasmall Cu NPs and $Cu_2O$ NPs from organo-copper(I) reagent. Reproduced with permission from ref. [67], Copyright 2017, American Chemical Society. (e) TEM, (f) AFM images of the ultrathin $Cu_2O$ nanosheets with 4 atomic thicknesses and the yellow line in f corresponds to the height profile in (g). Reproduced with permission from ref. [68], Copyright 2014, Elsevier.

It is difficult to control the morphology of $Cu_2O$ nanocrystals if the size drops below 10 nm because it tries to achieve a spherical structure to minimize the surface energy [65–72]. Also, unless a proper organic surfactant is added, the sub-10 nm nanocrystals undergo aggregation, forming bigger spherical nanocrystals [45]. O'Brien et al. synthesized highly uniform monodisperse Cu nanocrystals by reacting copper acetate with oleic acid and trioctylamine, while heated at 180 °C [65]. Upon further oxidation in the presence of air and ligand protection, those Cu nanocrystals were systematically converted to highly crystalline 6 nm $Cu_2O$ nanocrystals (Fig. 11a). A thin layer of CuO formed at the nanocrystal-ligand interface because of over-oxidation. The corresponding selected area electron diffraction patterns (Fig. 11b) confirmed the cubic crystal structure of $Cu_2O$. Oleic acid stabilized these nanocrystals in a solution, which prevented aggregation. A similar procedure was also reported to synthesize hollow and solid $Cu_2O$ nanocrystals of 8 nm and 14 nm in size, respectively [70]. Further expansion of this idea may lead to many reversible catalytic interconversions of the ligands protecting Cu and $Cu_2O$ nanocrystals in the presence of $O_2$ as an oxidant and $H_2$ as a reductant (Fig. 11d) [67]. During these interconversions, the used bidentate ligands strongly coordinate both Cu and $Cu_2O$ oxidized NPs surface by forming strong covalent bonds, which prevents them from aggregation and size distortion. One-dimensional (1D) nanomaterials such as nanowires are also highly attractive building blocks for photocatalysis because of the inherent anisotropy, which improves charge separation and photocatalytic efficiency [7,68]. $Cu_2O$ nanowires were also developed by different wet-chemical methods such as the decomposition of the metal-ligand complex route or the use of polyethylene glycol (PEG; Mw 20000) and hydrazine as the reducing agent at room temperature [71]. Most interestingly, Li et al. developed an easy approach to synthesizing bulk quantities of single crystalline $Cu_2O$ nanowires with tunable diameter and length. Uniform $Cu_2O$ nanowires were prepared through the reduction of cupric acetate with o-anisidine, pyrrole, or 2,5-dimethoxyaniline as the reductant in dilute aqueous solutions under hydrothermal conditions, as shown in Fig. 11c [66]. The length of the $Cu_2O$ nanowires varied from tens of micrometers to more than one hundred micrometers.

Two-dimensional (2D) materials such nanosheets are also very interesting due to the quick separation of electron–hole pairs from bulk to surface as well as for their large surface area and efficient light harvesting properties [68]. Xie et al. synthesized an ultrathin $Cu_2O$ nanosheet with 4 atomic thicknesses by decomposing lamellar $Cu_2O$-oleate complex intermediate microplates, which are formed by periodically stacking organic molecules and inorganic layers, at a temperature of 300 °C for 8 min in air. The TEM image in Fig. 11e shows an almost transparent feature, and the clear Tyndall effect indicates the ultrathin thickness of such $Cu_2O$ nanosheets, further verified by their atomic force microscopy (AFM) images, presented in Fig. 11f. The thickness of the nanosheets was around 0.62 nm (Fig. 11g). All the above-mentioned case studies provide not only a clear stepwise development of the synthesis of diverse well-defined $Cu_2O$ nanostructures but also a good opportunity to understand the growth mechanism and role of different ions, solution pH, and surfactant concentration. The remaining synthetic challenges in this field can be also overcome on the basis of these findings.

*2.3 Methodologies for removing capping agents.*

In the previous section, we discussed the role that the different surfactants and their concentrations paly in controlling both the shape and size of $Cu_2O$ nanocrystals. Common surfactants such as sodium dodecyl sulfate, trisodium citrate, cetyltrimethylammonium bromide, oleic acid, oleic amine, hexadecylamine, polyvinylpyrrolidone and polyethylene glycol are majorly used for the synthesis of well-defined $Cu_2O$ nanocrystals with the control of size. These surfactants not only stabilize the crystal plane but also prevent any agglomeration during the synthesis and the isolation process of the nanocrystals. After the isolation, the nanocrystals are employed in different catalytic processes in order to evaluate their facet-dependent properties. Different crystal facets have different atomic arrangements of $Cu^{+1}$ and $O^{-2}$ ions, resulting in different $Cu^{+1}$ ion surface atomic densities, which may lead to different catalytic activity [44]. If the surfactant molecules are still present on these facets, they can block the catalytically active sites and hinder the access of reactant molecules to the reactive sites [73–75]. Moreover, under particular reaction conditions, surfactants may decompose and contribute to the formation of hydrogen or carbon-based products, thus affecting the proper evaluation of the photocatalytic activity and product selectivity. This aspect is particularly delicate for $CO_2$ photoreduction and will be discussed in more detail in section *4.4*. Therefore, removing the capping agents from the crystal facets must precede any comparison of the facet-dependent catalytic activity. Thus, the surfactant molecules present on the crystal facet can act as poison in the catalytic reaction.

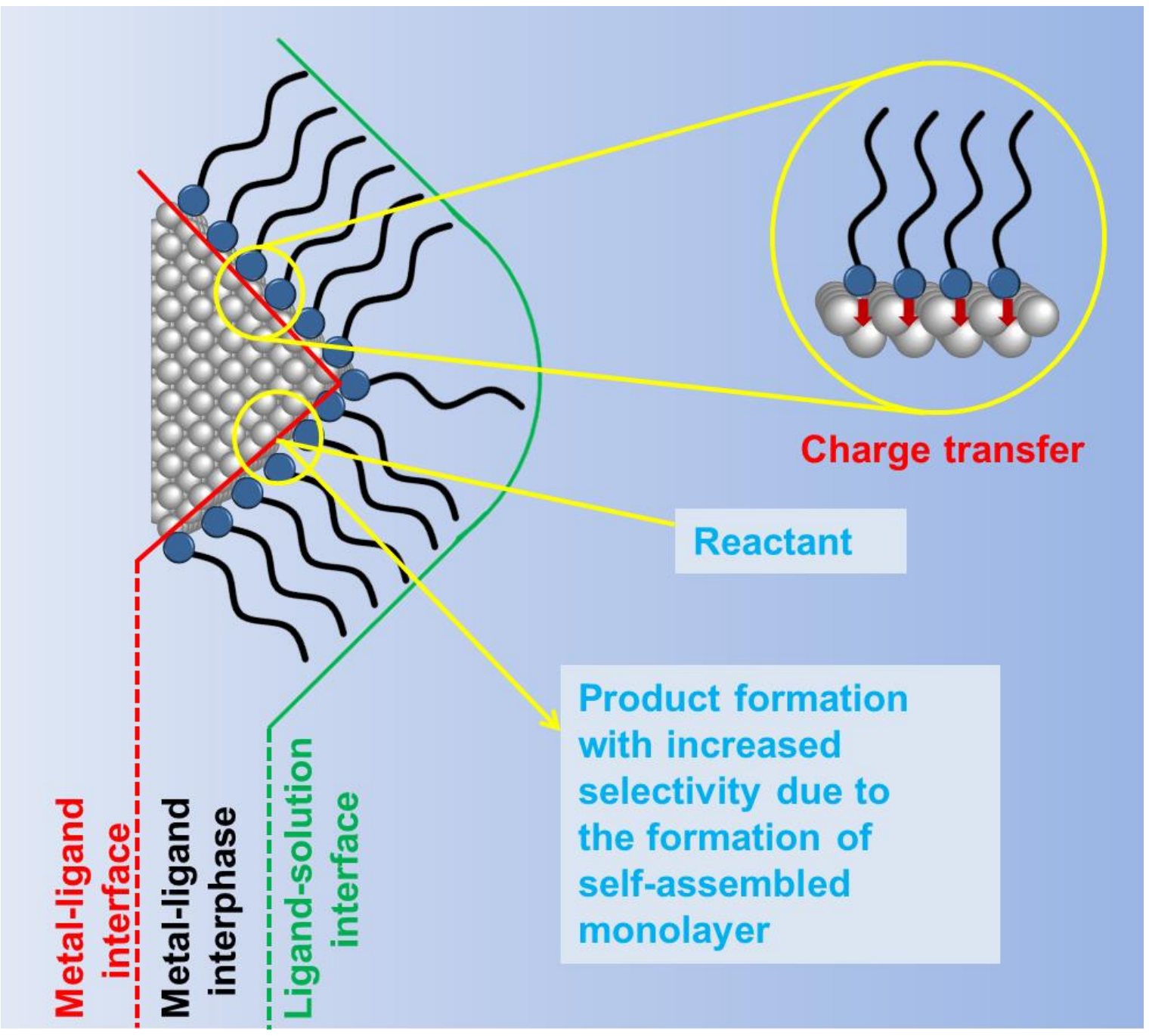

**Figure 12.** Schematic diagram showing the ligand orientation around a nanocrystal surface from a microscopic point of view. Influence of ligands on the nanocrystals surface and on product selectivity. Adapted with permission from ref. [73], Copyright 2016, MDPI.

It is important to understand the interface between the capping agent and the crystal facet on a nanoscale level before moving on to the removal section [73]. There are mainly two different interfaces such as 1) the metal-ligand interface and 2) the ligand-solution interface, as shown in Fig. 12. The intermediate transition zone known as the metal-ligand interphase can have more versatile complex structures depending upon the molecular weight, structure, and functional group of the surfactant molecules. There is a lack of literature that would evaluate the impact of different surfactant molecules on the activity and selectivity for $Cu_2O$ nanocrystals of different shapes. Most studies have been done on Pt, Au, Pd, and Ag metal nanocrystals [75]. At the metal-ligand interface, two major phenomena can occur, both benefiting the catalytic reaction carried out on a metal nanocrystal. Firstly, the discrete molecular orbitals of the ligands' functional groups can strongly interact with the metal atoms present on the nanocrystal surface and induce a charge transfer from the ligand to the metal atom. This effect can modify the local charge density and facilitate any of the steps involved in the catalytic reaction such as adsorption of the reactants, bond-breaking, bond-formation, and desorption of the product molecules, making the reaction more or less selectively [73–75]. For example, electron donation from an electron-rich capping agent to the surface metal atoms has been used to control the product selectivity of the reaction, by favoring the adsorption of electron deficient substrates and promoting the desorption of electron-rich product molecules from the surface. A second most important effect is the formation of a self-assembled monolayer (SAM) around the nanocrystal surface. Due to the strong affinity of the thiols group present in an alkane thiol, the surfactant adsorbs more strongly with the head group to the metal atoms, while the hydrocarbon tail remains oriented at a well-defined angle with respect to the surface forming the SAM structure. Such an ordered morphology of the SAM can induce a steric hindrance, which can limit the mobility of differently sized reactants and force them to rearrange the adsorption geometry close to the active site, which enables controlling the selectivity or more importantly the stereoselectivity of the products.

Apart from the above-mentioned promotional effects of the capping agent, it is still necessary to completely remove the surfactant molecules from the $Cu_2O$ nanocrystals surface in order to compare their facet-dependent catalytic activity in different photocatalytic reactions as these capping agents can reduce the photocatalytic activity by neutralizing the photogenerated electrons or holes via a charge transfer. Completely "clean surface" in the solution of nanocrystals cannot be truly obtained, but it can be considered to be free from long-chain sticky surfactant molecules [74,75]. After the removal of capping agents, the clean surface is stabilized by small molecules such as solvent molecules or solute ions, which are easily substituted by reactants during catalytic reactions.

There are different, well-established methods to remove the capping agents from the nanocrystal surface [73–75]. The most common procedure consists of consecutive sonication steps involving large amounts of solvents and centrifugation to collect the nanoparticles. For example, surfactants such as sodium dodecyl sulfate and trisodium citrate can be easily removed by washing with 1:1 water/ethanol mixture for several times followed by centrifugation [76,77]. Therefore, synthetic procedures that use SDS or trisodium citrate as a surfactant for the synthesis of $Cu_2O$ nanocrystals are very promising since clean-surface can be obtained by a simple washing procedure. Other surfactants such as CTAB, oleic acid, oleic amine, hexadecylamine, polyvinylpyrrolidone, and polyethylene glycol can also be removed from the nanoparticle surface, partially using extensive washing steps, which can lead to the deformation of crystal facet or leaching. Different strategies have been developed to deal with this problem. One of the ways is to decompose the capping agents into small molecular fragments by heat or light, followed by their removal by means of a gas flow or solvent [78]. This method includes high-temperature thermal annealing and UV-ozone (UVO) irradiation. Surfactants with higher molecular weight such as PVP and PEG can be also removed following this procedure [73–75]. One major drawback of this procedure is the partial decomposition of the organic surfactant during the thermal treatment or UVO irradiation with the consequent generation of coke and other species, which can deactivate the catalyst facet. Another strategy makes use of the addition of pure acetic acid to amine-capped nanoparticles causing the protonation of amino groups, which weakens their interaction with nanocrystal surface [73–75]. This accelerates their detachment during the solvent-washing procedure. Surfactants such as oleylamine, hexadecylamine, and oleic acid can be washed off this way. Other strategies are based on adding excessive small molecules (such as 1-butylamine) to displace long-chain hydrocarbons organic surfactants (such as oleylamine, PVP), which also have a stronger capping ability. Due to the much lower boiling point and mass, smaller molecules can be easily removed by extensive washing or vacuum evaporation [73–75].

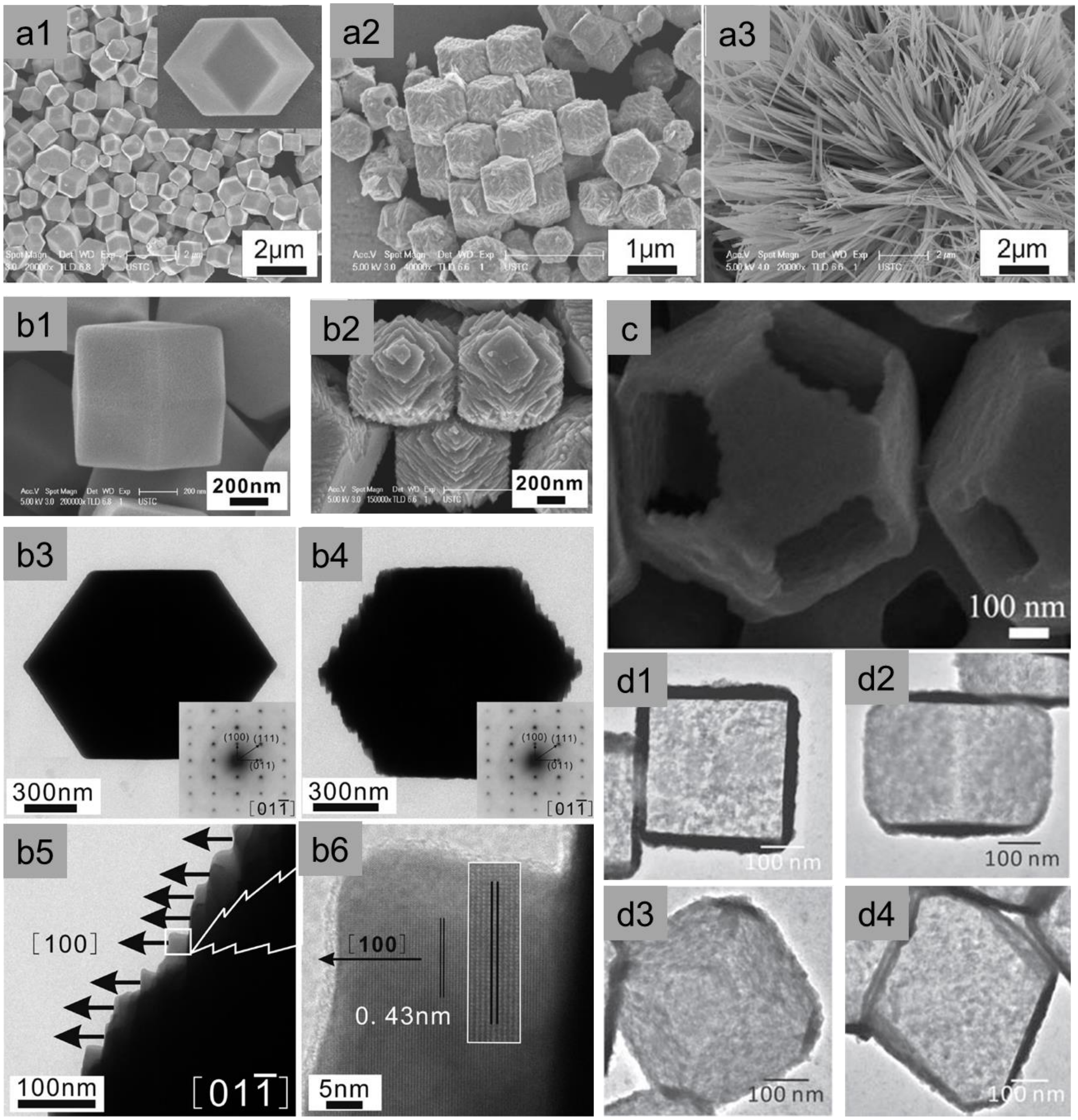

**Figure 13.** SEM image of as-synthesized (a1) $Cu_2O$ rhombic dodecahedral (RD). After exposing these microcrystals in aqueous ammonia solution for (a2) 3 min and (a3) 40 min. Reproduced with permission from ref. [79], Copyright 2011, American Chemical Society. (b1) SEM and (b3) TEM image of as-synthesized $Cu_2O$ RD. After exposing these microcrystals in aqueous acetic acid solution for time 150 min (b2) SEM, (b4) TEM and (b5, b6) corresponding HRTEM images with the image of lattice fringes. Reproduced with permission from ref. [80], Copyright 2011, American Chemical Society. (c) High-magnification FESEM image of a $Cu_2O$ nanoframe obtained after exposing in air at 20 °C for 16 days. Reproduced with permission from ref. [81], Copyright 2011, Wiley-VCH. TEM images of a single cubic and octahedral $Cu_2S$ nanocage

viewed over its (d1, d3) {100} face and (d2, d4) {110} edge respectively. Reproduced with permission from ref. [82], Copyright 2011, Wiley-VCH.

After discussing general approaches to removing the surfactant from the surface of different inorganic nanocrystals, we will now shift the focus onto those procedures that are specifically suitable for $Cu_2O$ nanocrystals. It is highly desirable that during the surfactant-removal process, the shape, size, composition, crystallinity, and crystal facet of a $Cu_2O$ nanocrystal are preserved. If we change any one of these parameters, the results of studying the facet-dependent properties will be erroneous. It is also crucial to remember that $Cu^{+1}$ ions in $Cu_2O$ can undergo a disproportionation reaction under different chemical environments forming $Cu^{+2}$ or $Cu^0$ states, which may lead to the chemical decomposition of the crystal or structural modification. Different inorganic ions or molecules can also cause a serious damage to well-defined $Cu_2O$ nanocrystals, and these must be identified before conducting the surfactant removal process or pursuing any photocatalytic reactions.

When $Cu_2O$ microcrystals are exposed to an aqueous solution of ammonia (pH~11.4) for 3 or 40 min under stirring condition, it will bring a certain change to the shape, size and composition, as demostrated in Fig. 13a1–a3 [79]. After 3 min, rhombic dodecahedral $Cu_2O$ microcrystals undergo an etching process which transforms a smooth {110} surface into a rough and stepped square facets. HRTEM images confirmed that the stepped square facets belongs to the $Cu_2O$ (100) crystal plane. When the reaction time reaches 40 min, $Cu_2O$ rhombic dodecahedral shapes were completely converted into $Cu(OH)_2$ nanobelts (Fig. 13a3), as confirmed by PXRD and XPS. It is important to note that in an aqueous NaOH solution with the same pH value such etching processes of $Cu_2O$ nanocrystals cannot be observed [79]. Therefore, it can be postulated that the etching process is initiated by the coordination of surface $Cu^{+1}$ ions present in $Cu_2O$ microcrystals with $NH_3$ to form $[Cu(NH_3)_4]^{+1}$ species. They undergo an oxidation process by dissolved $O_2$ to produce $[Cu(NH_3)_4]^{+2}$, which then slowly precipitates in an aqueous ammonia solution to produce $Cu(OH)_2$. Controlled experiments proved that the stability of different low index $Cu_2O$ crystal planes in the aqueous ammonia solution followed the order {100}> {111} > {110}, suggesting that the cubes are more stable than rhombic dodecahedral morphology in aqueous $NH_3$ solution [79]. A similar stability trend can also be observed when $Cu_2O$ microcrystals are exposed to an acetic acid solution (pH=3.5) under stirring condition [80]. Oxidative etching of octahedral and rhombic dodecahedral $Cu_2O$ microcrystals in the weak acid solution leads to the formation of stepped layers of square facets exposing {100} planes (Fig. 13b1–b6), evidencing that the $Cu_2O$ {100} planes are more stable than the $Cu_2O$ {111} and {110} planes. But, in this case, surface chemical composition remains similar to that of $Cu_2O$.

Further, a truncated octahedral $Cu_2O$ microcrystals has eight hexagonal {111} faces and six {100} faces. When these microcrystals are exposed to air in the presence of PVP, selective etching leads to the formation of a single crystalline $Cu_2O$ nanoframes surrounded only by {111} faces, whereas all {100} faces are etched away completely (Fig. 13c) [81]. This contradictory event can be explained as follows: PVP acts as a capping agent, and preferentially adsorbs on the {111} facets of the $Cu_2O$ crystals, which prevents the {111} planes from etching; therefore, {100} facets completely vanish, leading to the formation of hollow structures. Any sulfide anions are also detrimental to $Cu_2O$ as they easily replace the oxide anions from the crystal lattice, which leads to the formation of $Cu_2S$ crystals [82]. A single crystalline cubic and a truncated octahedral $Cu_2S$ nanocage have been prepared with ultrathin walls through controlled sulfidation of the corresponding $Cu_2O$ crystals, and the acid etching was applied to remove the interior $Cu_2O$ portions (Fig. 13d1–d4). This suggests that any sulfide species should always keep away from $Cu_2O$ nanocrystals during the surfactant removal, surface modification or photocatalytic reaction.

Not only sulfide but also thiosulfate anions ($S_2O_3^{-2}$) can induce a strong coordinating etching process on $Cu_2O$ nanocrystals. Different metal (M = Mn, Fe, Co, Ni, Zn) hydroxide nanocages can be obtained by reacting a certain amount of $Cu_2O$ templates with $MCl_2$ $yH_2O$ in an ethanol/water mixed solvent in the presence of PVP ($M_w$ = 30,000) and $Na_2S_2O_3$ [83]. The formation mechanism can be explained as follows: as per Pearson's hard and soft acid−base (HSAB) principle *"soft base form strong complexes with soft acid whereas hard base prefer hard acid"* [83]. Thus thiosulfate anion (soft base) forms stronger bond with $Cu^{+1}$ cations (soft base) in $Cu_2O$ and replace the lattice oxide anions (hard base). As a result, coordinating etching of $Cu_2O$ occurs by forming a soluble $[Cu_2(S_2O_3)_x]^{2-2x}$ complex, which is followed by a synchronous complex precipitation process of $M(OH)_2$ around the $Cu_2O$ template surface, leading to the formation of metal hydroxide nanocages. This process is termed as "coordinating etching and precipitating" (CEP) [83]. Similar to acetic acid, dilute HCl can also induce facet-selective etching of $Cu_2O$ nanocrystals, leading to the formation of diverse single crystalline cubic $Cu_2O$ nanoframes [84,85]. Therefore, it is essential to remember that acetic acid, hydrochloric acid, ammonia, aqueous solution exposed to air, sulfide, and thiosulfate anions can cause serious damage to the shape and composition of the $Cu_2O$ nanocrystal. They should not be used during the removal of capping agents or even in photocatalytic reactions. It should be noted that SDS or trisodium citrate would be a better choice during the synthesis of $Cu_2O$ nanocrystals as they can be easily removed by consecutive washing with water/ethanol mixture.

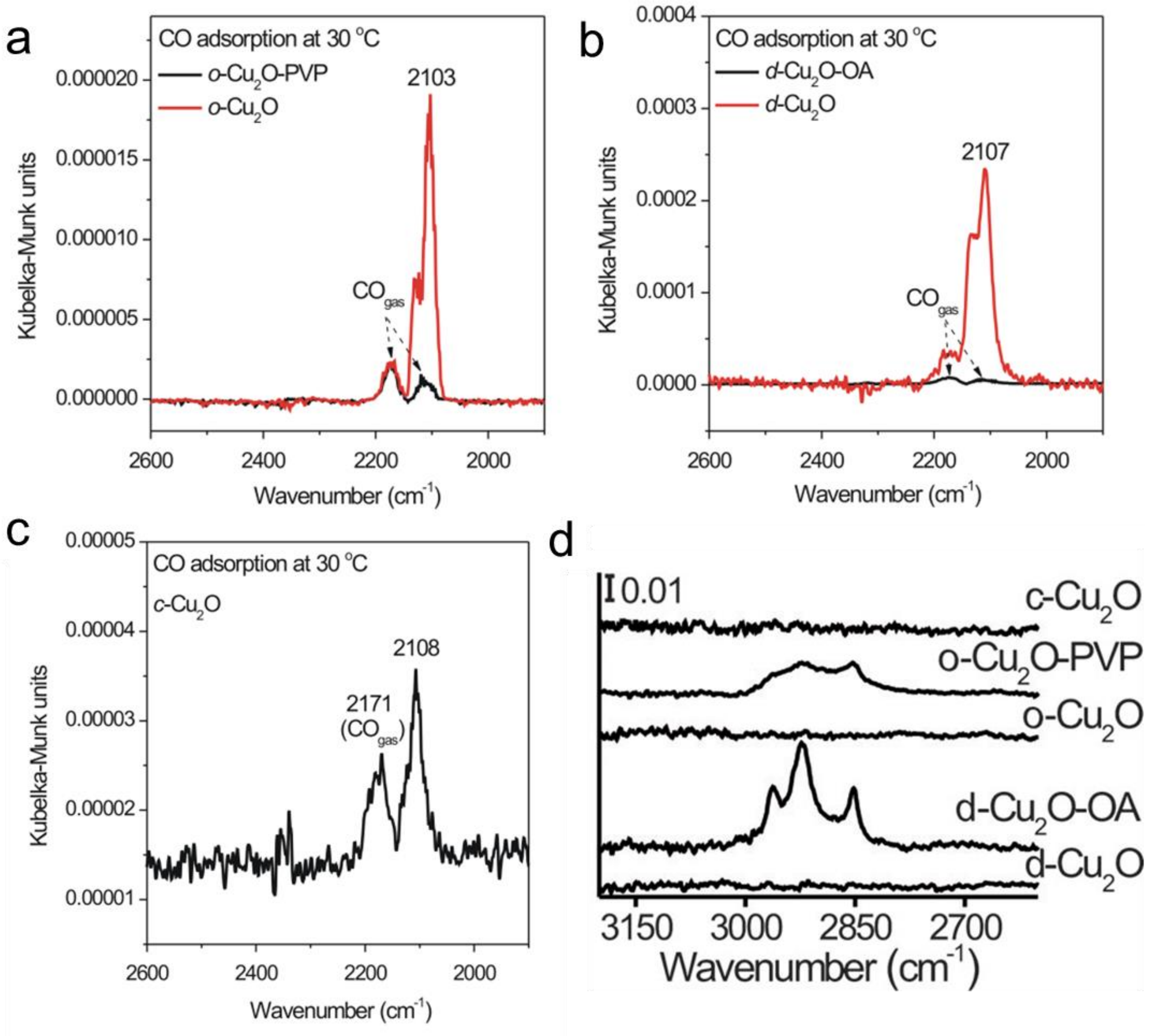

**Figure 14.** DRIFTS spectra of CO chemisorption at 30 °C on (a) $Cu_2O$ octahedra capped with PVP (o-$Cu_2O$-PVP) and capping ligand-free $Cu_2O$ octahedra (o-$Cu_2O$), (b) $Cu_2O$ rhombic dodecahedra capped with OA (d-$Cu_2O$-OA) and capping ligand-free $Cu_2O$ rhombic dodecahedra (d-$Cu_2O$), and (c) capping ligand-free $Cu_2O$ cubes (c-$Cu_2O$). (d) FTIR spectra of the corresponding nanocrystals. Reproduced with permission from ref. [86], Copyright 2014, Wiley-VCH.

Regarding PVP, PEG, oleylamine, oleic acid, and hexadecylamine, extra care is needed during the removal steps. The successful removal of the organic surfactant can be than confirmed by FTIR, XPS, and a CO-chemisorption study. Huang et al. used PVP and oleic acid (OA) as capping ligands to prepare octahedral and rhombic-dodecahedral $Cu_2O$ microcrystals, respectively, whereas no capping ligands were used for the cubic shape [86]. PVP and OA were completely removed from the $Cu_2O$ crystal surface by placing them in a U-shaped quartz microreactor, followed by purging in a stream of $C_3H_6$, $O_2$ and $N_2$ and heating to the desirable temperature to carry out a controlled oxidation treatment. Oxygen led to the oxidation of capping ligands at low temperature, whereas the coexistence of $C_3H_6$, $O_2$ and $N_2$ in the atmosphere cooperatively prevented the $Cu_2O$ surface from over-oxidation. A clear CO chemisorption peak could be observed after the removal of the surfactant but not on the capped $Cu_2O$ microcrystals (Fig. 14a–c). FTIR (Fig. 14d), PXRD, and XPS also confirmed the successful removal of PVP and OA from the $Cu_2O$ microcrystal surface, without changing the morphologies, surface compositions, and structures. In conclusion, with respect to the different types of surfactants used in the synthesis of $Cu_2O$ nanocrystals, we must conduct the surfactant-removal steps very carefully, keeping in mind all the previous discussion, so that we do not damage the well-defined nanocrystals before using them in photocatalytic applications. Either multiple washing with a solvent or heat and light treatment under appropriate conditions can be more suitable for achieving this target.

## 3. Facet-dependent properties.

Well-defined nanocrystals with different crystal facets (i.e. anisotropic facets) show different atomic arrangements. For this reason, they exhibit unique facet-dependent properties such as molecular adsorption, anisotropic redox reaction sites, surface electronic structures, and optical properties. All these facet-dependent properties can have a profound impact on photocatalytic properties; therefore, they will be discussed in detail in the following section [17].

*3.1 Selective deposition of co-catalyst.*

Semiconductor nanocrystals enclosed by anisotropic facets have different electronic band structures (i.e., surface states), which promote the accumulation of the photogenerated electrons and holes on different facets for the reduction and oxidation reaction, respectively [15,17]. The presence of a suitable co-catalyst on these redox sites can further enhance the electron-hole separation process, resulting in an enhanced photocatalytic activity. The facet-selective adsorption behavior of surfactants has been deployed to achieve selective deposition of the co-catalyst on $Cu_2O$ nanocrystals with anisotropic facets.

To probe this phenomenon, Choi et al. electrochemically deposited micrometer-sized cubic, octahedral, and truncated octahedral $Cu_2O$ crystals on indium tin oxide (ITO) substrates [87]. In the absence of a SDS surfactant, after 15 s, the electrochemical gold deposition showed that gold nanoparticles formed on both {100} and {111} planes (Fig. 15a–c) regardless of the crystal shape, i.e., lack of selectivity was observed. When SDS was introduced into the reaction medium, gold nanocrystals deposition took place on all of the {100} planes belonging to the $Cu_2O$ cubic crystals (Fig. 15d), which confirmed that SDS did not adsorb strongly on the {100} planes. However, no gold nanocrystals were found on octahedral $Cu_2O$ crystal (Fig. 15e), which indicated that SDS molecules strongly adsorbed on the {111} planes, resulting in the effective inhibition of the nucleation of gold nanocrystals. More surprisingly, gold nanocrystals were selectively deposited only on the {100} planes of truncated octahedral $Cu_2O$ crystals, which contained both the {100} and the {111} planes (Fig. 15f), thus confirming the SDS molecules preferential adsorption on the {111} planes of the $Cu_2O$ crystal. Therefore, we can choose different surfactant molecules to be used in combination with diverse $Cu_2O$ morphologies in order to investigate their preferential absorption and obtain facet-selective deposition of co-catalysts, which will help to improve the activity and selectivity of the investigated photocatalytic reactions. Even in the absence of any reducing agents, $HAuCl_4$ can be reduced to Au nanocrystals in the presence of $Cu_2O$ nanocrystal at room temperature; the driving force for this in-situ reduction can be explained in terms of the standard reduction potential difference between $AuCl_4^-$/Au [0.93 V vs. standard hydrogen electrode (SHE)] and $Cu^{+2}$/$Cu^{+1}$ [0.15 V vs. SHE] pairs [88,89]. A similar galvanic displacement reaction can also be seen for $H_2PdCl_4$ and $H_2PtCl_6$ precursors. During this process, a certain degree of surface etching can occur in $Cu_2O$ crystals, which will be spontaneously covered by newly-formed metal nanocrystals. Du et al. showed that in the absence of any surfactants, Au nanocrystals were first selectively deposited only on all tips of an octahedral shape via in-situ reduction, and then they covered the edges when using an increased amount of a gold precursor solution. Further, increase of amount of Au caused random coverage of all facets, edge, and tips of the octahedral $Cu_2O$ nanocrystals, indicating the loss of selective deposition. The main reason behind the observed behavior is that different facets possess different surface energy that follows the order as: $\gamma_{(facets)} < \gamma_{(edges)} < \gamma_{(tips)}$. Thus, the selective deposition of metal nanocrystals on $Cu_2O$ nanocrystals with anisotropic facets can be controlled by utilizing preferential adsorption of additives or controlling the amount of added metal precursors under suitable reaction conditions. Such hybrid nanocatalyst efficiently reduce the photogenerated electron–hole recombination, which results in an enhanced photocatalytic reaction rate [15].

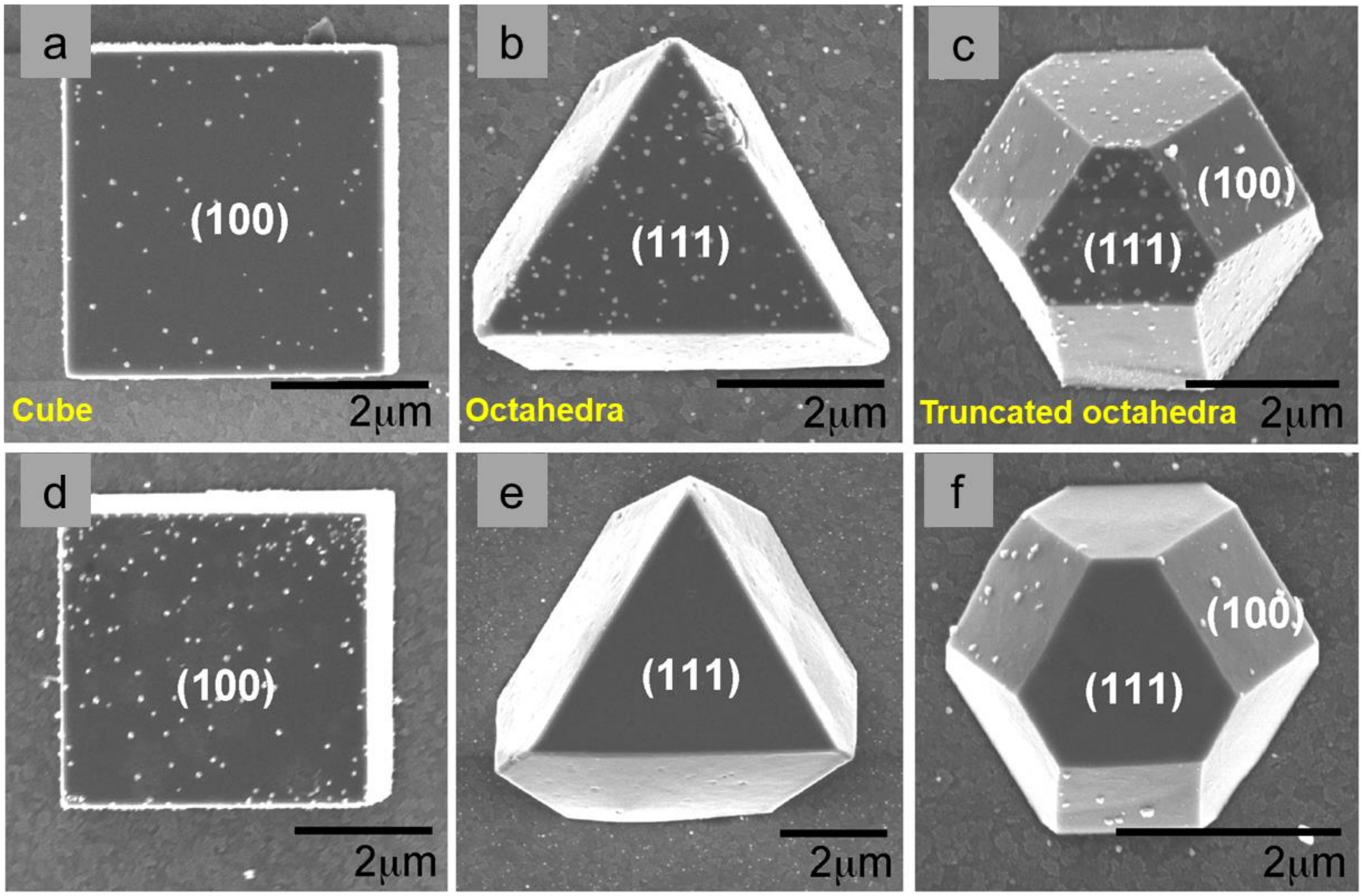

**Figure 15:** Surfactant assisted facet selective Au deposited on $Cu_2O$ crystals without (a–c) and with (d–g) sodium dodecyl sulfate. Reproduced with permission from ref. [87], Copyright 2009, American Chemical Society.

*3.2 Distribution of electrons and holes on different facets upon illumination.*

Efficient separation of photogenerated electrons and holes is one of the bottlenecks in achieving very high photocatalytic activity when using a semiconductor nanocrystal [15,17,26]. Photogenerated electrons should move to a reductive site to perform the reduction reaction, whereas holes need to reach oxidative sites to perform the oxidation reaction (Fig. 1a). For instance, nanospheres suffer from a high electron-hole recombination rate due to their isotropic geometry. Li et al. extensively used surface photovoltage (SPV) techniques based on Kelvin probe force microscopy (KPFM) and showed that the photogenerated charges can be separated effectively in a high-symmetry $Cu_2O$ photocatalyst by asymmetric light irradiation [90–94]. The holes and electrons were transferred to the illuminated and shadow regions, respectively, of a single photocatalytic particle. Further investigation in this regard will be needed to identify the highly efficient catalytic sites on $Cu_2O$ nanocrystals with anisotropic facets. Single-particle single-molecule fluorescence photocatalysis could be a possible tool for on-site observation of interfacial chemical reactions involving charge carriers and reactive oxygen species (ROS), such as singlet oxygen and the hydroxyl radical, generated by the photoexcitation of semiconductor nanocrystals [95,96]. Such studies confirmed that the presence of anisotropic facets can be beneficial to obtaining efficient spatial charge separation. Using the small sized $Cu_2O$ crystals with anisotropic facets remains a challenge. It could help to achieve a higher photocatalytic performance in different reactions, which results from their increased high reactive surface area.

*3.3 Optical properties.*

Before moving on to the photocatalytic application, it is very important to understand how different sizes and shapes of $Cu_2O$ nanocrystals affect their optical properties. Diverse synthetic strategies give us an opportunity to better understand the existence of facet-dependent optical properties in $Cu_2O$ nanocrystals, which is barely studied [97,98]. The absorption peak for $Cu_2O$ nanocrystals varies from 450 nm to 700 nm, depending upon the shape and size [99]. Generally, while increasing the volume of a spherical $Cu_2O$ nanocrystal, the absorption peak will red-shift and broaden systematically [99]. However, surprisingly, $Cu_2O$ nanocrystals with different facets of similar volume can have different absorption peak positions in the same solvent, which is a purely facet-dependent optical property [100–103]. Figure 16a compares the absorption peak position of five different comparable volumes of octahedral and cubic $Cu_2O$ samples in absolute ethanol. Interestingly, nanocubes show a more significant absorption band redshift than octahedra of similar volumes. Moreover, the absorption band separation between the corresponding cubes and octahedra is 13–16 nm, and is independent of particle volume [100]. This phenomenon is directly related to the exposed crystal facet of the $Cu_2O$ nanocrystals. The observed facet-dependent optical effects to the fact that a thin $Cu_2O$ surface layer (~less than 1.5 nm) has different band structures and different degrees of band bending with respect to the individual facets of $Cu_2O$ [103]. The thicknesses of these surface layers were determined through density functional theory (DFT) calculations, which were approximately 6.2, 11.7, and 4.5 Å for the {111}, {100}, and {110} facets of $Cu_2O$ respectively. For the three surface atomic layers of (111), (100), and (110) planes, the density of state (DOS) plots resembled those of a metal, a semimetal, and a semiconductor, respectively. Thus, one can treat these thin surface layers of $Cu_2O$ as being composed of different materials with different refractive indices. As the refractive index ($n$) of the medium is directly related to its dielectric constant via $\varepsilon_m = n^2$, a thin shell material with a high dielectric constant causes large redshifts. Figure 16b represents the photoluminescence spectra for three differently sized $Cu_2O$ RD nanocrystals showing that the intensities are highest for the smallest nanocrystal [102]. Also $Cu_2O$ nanocrystals with three different shapes show different colors (Fig. 16c) due to their different band gap energy [102]. Thus, it can be concluded that crystal facet engineering can help to tune the band gap of a semiconductor.

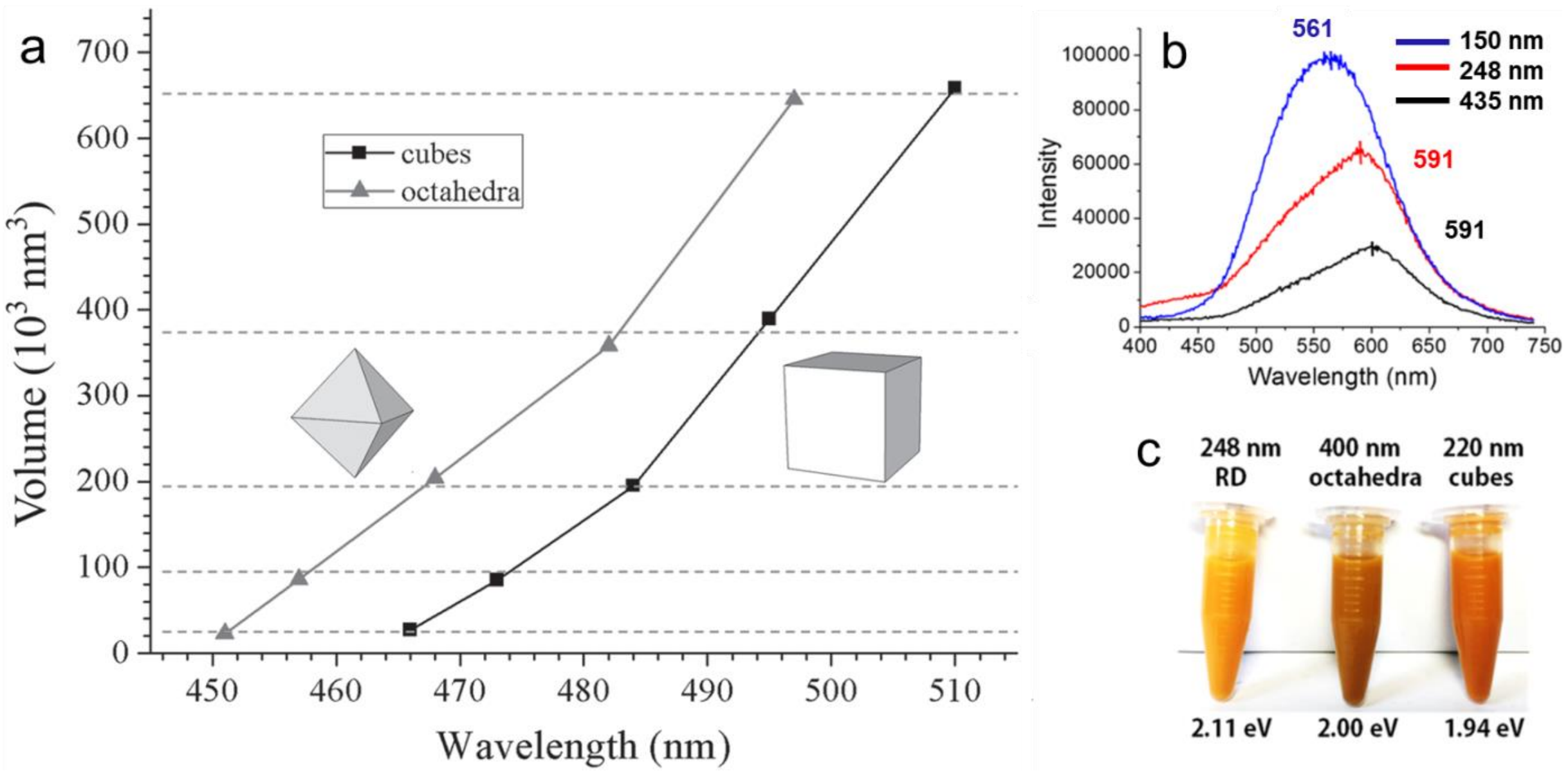

**Figure 16:** (a) A comparison plot for the variation of volume for $Cu_2O$ octahedra and cubes with respect to their absorption peak positions. Reproduced with permission from ref. [100], Copyright 2016, Wiley-VCH. (b) Photoluminescence spectra of different sized $Cu_2O$ RD nanocrystals. Excitation wavelength used was 380 nm. Reproduced with permission from ref. [102], Copyright 2018, American Chemical Society. (c) Color comparison of three different shaped $Cu_2O$ nanocrystals, confirming visually observable optical facet effects. Reproduced with permission from ref. [102], Copyright 2018, American Chemical Society.

Most interestingly, these thin $Cu_2O$ surface layers exposing the (111), (100), and (110) planes with different dielectric constants not only influence the absorbance peak position of pure $Cu_2O$ nanocrystals but can also cause a large redshift in the localized surface plasmon resonance (LSPR) peak of noble nanomaterials [104–107]. Huang group synthesized Au–$Cu_2O$ core-shell nanocrystal with octahedral, cuboctahedral, and cubic morphology with high uniformity in size using 50 nm octahedral Au nanocrystals as a core, which showed a LSPR peak at 550 nm (Fig. 17a) [104]. The colloidal solution color gradually changed from a dark orange color to greenish blue, while the average particle size decreased (Fig. 17b). Surprisingly, the UV−Vis absorption spectra of the Au–$Cu_2O$ core–shell nanocubes, cuboctahedra, and octahedra showed a fixed

LSPR peak position respectively at 752, 768, and 778 nm, though there was a systematic change in the $Cu_2O$ shell thickness for each shapes. It should be noted that the SPR peak positions are also dependent on the exposed surface facets of the $Cu_2O$ shell. Despite the different shell thicknesses and overall sizes, the LSPR peak of the Au core was more redshifted for the cubic morphology (778 nm) than for the octahedral shape (752 nm), whereas cuboctahedral felt in the middle (768 nm) as they both had the {100} and {111} facets. This can be due to the different $Cu_2O$ crystal facets having a thin surface layer with different dielectric constant. Therefore, when a plasmonic wave propagates from the metal core to the $Cu_2O$ shell, it sees the surface layer with different dielectric constants and responds differently to this environmental change, shifting the LSPR peak position accordingly. To further verify this rare optical phenomenon, diverse types of core-shell nanocrystals were also synthesised and studied [105–107]. Interestingly, different surface oxidation states or incorporation of various amino acids into the crystal lattice of $Cu_2O$ nanocrystals can also lead to different optical and band gap energies [108,109]. The unique optical properties of such hybrid nanostructure may also be beneficial for different photocatalytic and photothermal applications as they strongly absorb in the whole solar spectrum.

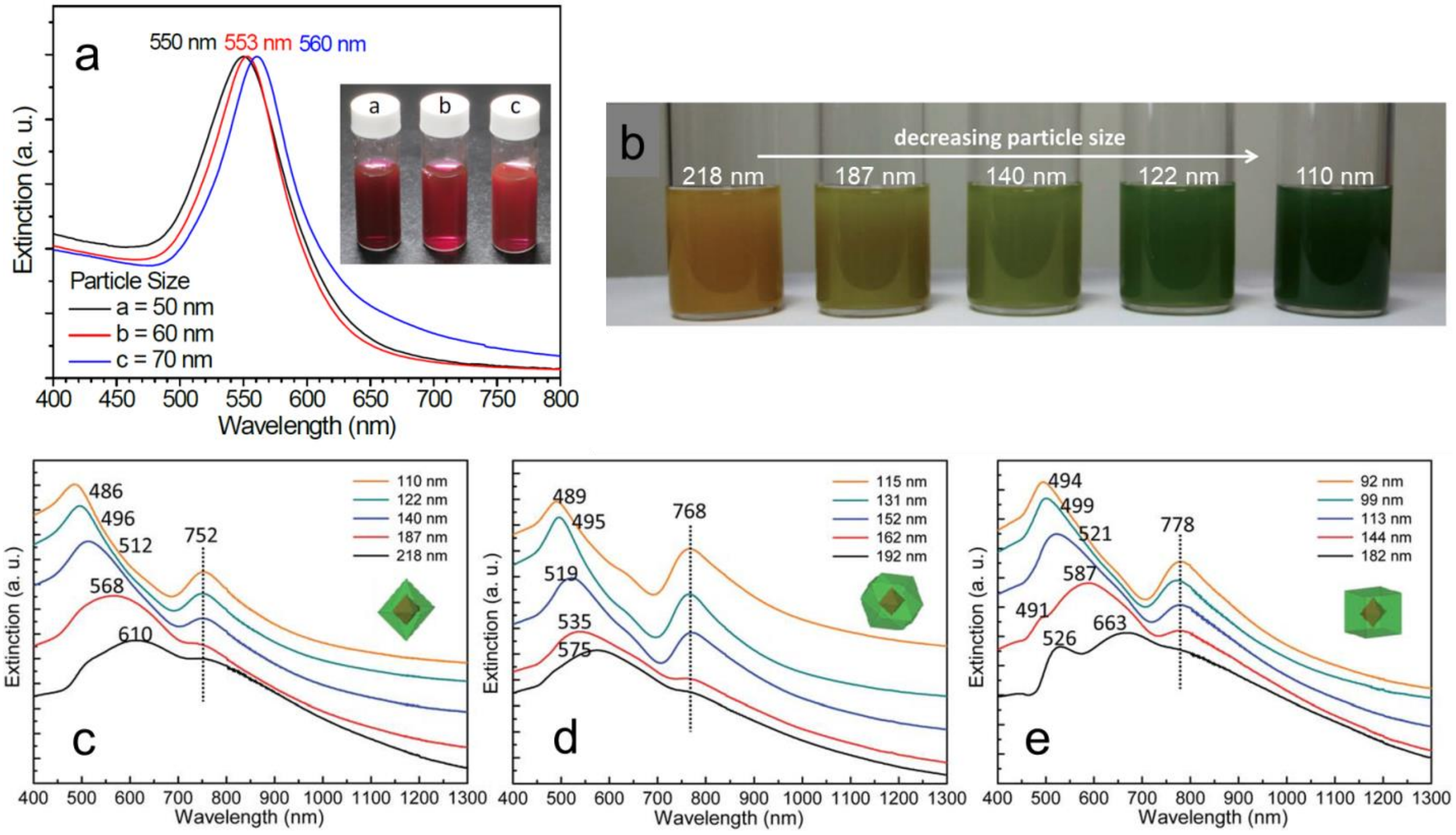

**Figure 17:** (a) UV–vis absorption spectra of the different sized octahedral Au nanocrystal. (b) Color comparison of different sized octahedral Au–$Cu_2O$ core–shell nanocrystal. UV–vis absorption spectra of the five different sized Au–$Cu_2O$ core–shell nanocrystal with (c) octahedral, (d) cuboctahedral and (e) cubic shapes having 50 nm octahedral Au core inside it. Reproduced with permission from ref. [104], Copyright 2014, Royal Society of Chemistry.

## 4. Photocatalysis with well-defined $Cu_2O$ nanocrystals

In this section, we will discuss the proper ways of representing the common parameters that are directly related to photocatalysis. Photocatalytic applications of well-defined hybrid $Cu_2O$ nanocrystals have been studied for different organic reactions, $H_2$ evolution and the $CO_2$ reduction reaction. The photo-instability of $Cu_2O$ nanocrystals remains a key challenge to overcome during catalysis. Therefore, different strategies to solve this problem have been emphasized and accompanied with a proper case study.

### *4.1 Figures of merits in photocatalysis.*

Before starting the discussion about well-defined $Cu_2O$ nanocrystal-mediated photocatalytic processes, it is necessary to introduce the figures of merit commonly used to assess the photocatalytic performance of a newly designed photocatalyst with other reported benchmark photocatalysts. It is very common to notice that the absence of this crucial information in research papers makes it difficult to compare the photocatalytic efficiency, and also reproduce it, if necessary. Therefore, careful calculation and proper use of the most important figures of merit in research papers should be provided in order to help the research community in this field to facilitate the development of a highly efficient photocatalyst for different photocatalytic reactions [110–117]. In this chapter, we will cover the following figures of merit: (a) turnover frequency, (b) specific surface area, and (c) apparent quantum yield and solar quantum yield.

*(a) Turnover frequency (TOF)*: TOF is a traditional figure of merit employed in heterogeneous catalysis providing metrics for the number of molecules produced with respect to the amount of employed catalyst in a given chemical reaction in the unit time. Comparison of TOF for the given photocatalytic process can be useful for understanding how active the designed photocatalyst is, compared to others. TOF can be expressed as follows:

$$TOF\ (mol_{product} mol_{catalyst}^{-1} time^{-1}) = \frac{mol\ of\ product}{mol_{catalyst} \times time\ of\ reaction}$$

If a hybrid photocatalyst is made of two or more different components, it will be difficult to calculate the molar mass of such a material. In that situation, for simplicity, $mol_{catalyst}$ is often replaced with $g_{catalayst}$ and the TOF will be expressed as $mol_{product} g_{catalyst}^{-1} time^{-1}$. This way is very often used to compare the performance for photocatalytic $CO_2$ reduction and $H_2$ evolution reactions even if the data provide essential technical information and do not allow a proper comparison of materials with different surface area [110,116].

If we can identify and quantify the fully exposed active sites in a hybrid catalyst that solely carry out the catalytic reaction, than TOF can be expressed as follows

$$TOF\ (mol_{product} mol_{active\ site}^{-1} time^{-1}) = \frac{mol\ of\ product}{mol_{active\ site} \times time\ of\ reaction}$$

*(b)* *Specific activity per surface area*: Catalytic reactions mainly occur on the exposed catalyst surface. Thus, catalytic reaction rates (or TOF values) are directly proportional to the available surface area of the catalyst. Therefore, surface area is an important parameter, which needs to be reported during the catalytic rate measurement.

Generally, specific surface area is defined as the total surface area of a nanocatalyst per unit of mass. Its unit is expressed as $m^2\ g^{-1}$ and it must be measured before making any comparison of the facet-dependent photocatalytic reactivity of differently shaped nanocrystals. In a photocatalytic reaction, the amount of differently shaped nanocatalysts must have the same surface area in order to make a fare comparison of facet-dependent photocatalytic reactions. This condition is not often met, and, therefore, one must report the photocatalytic activity normalized over the surface area, which can be termed as specific activity per surface area [7].

$$Specific\ activity\ per\ surface\ area = \frac{mol\ of\ product}{surface\ area\ of\ the\ used\ catalyst \times time}$$

Such practice provides greater insight into the photocatalytic activity of the investigated well-defined nanocrystals. If the synthesized nanocrystals possess a highly uniform shape and size, the surface area can be simply calculated from a particle size distribution plot and also with respect to the geometry of that particular nanostructure. However, if the nanocrystals are not uniform, the specific surface area can be measured by adsorption-desorption isotherm adopting the Brunauer–Emmett–Teller (BET) method. This method is a more precise way to determine the specific surface area of a nanocrystal as photocatalysts often have a certain degree of mesoporosity, which may increase the final surface area of the material, but this cannot be considered when using mere geometrical calculation. Finally, determination of the surface area of supported nanostructure is far from being trivial.

*(c)* *Apparent quantum yield (AQY)*: Generally, quantum yield (QY) for a photochemical reaction is defined as the amount of reactants consumed or products formed per photons of light absorbed by the compound responsible for product formation [116]. If a specific monochromatic wavelength is used than QY becomes AQY. It helps to understand the wavelength-dependent photocatalytic behavior of a photocatalyst and to compare its efficiency with that of other materials, independently from the adopted experimental conditions. To evaluate it, the experiments must be carried out under different monochromatic light irradiation [118]. Photon flux must be always reported during the demonstration of experimental procedures.

First, reaction rate (k) needs to be expressed in $\left(\frac{molecules}{s}\right)$ unit for each reaction.

Then, the incident photon flux (Φ) can be calculated as

$$\Phi = \frac{\lambda}{hc} H \; [s^{-1} m^{-2}] \qquad (1)$$

where $\lambda$ is the wavelength of incident monochormatic light, $h = 6.626 \times 10^{34}\, J.s$, $c = 3 \times 10^{8}\, m.s^{-1}$ is the speed of light, and $H$ is the light intensity of monochromatic irradiation. By combining equations (1) and (2), the AQY cab be calculated as

$$AQY\% \left(\frac{molecules}{photon}\right) = \frac{k}{\mathrm{A} \times \Phi} \times 100$$

$where\ \mathrm{A}\ (\text{Avogadro number}) = 6.023 \times 10^{23}\, mol^{-1}$

Consistently, AQY must be reported with indication of the wavelength at which it was calculated and, obviously, comparison between AQYs' samples must be conducted at an identical wavelength.

Similarly, the solar AQY, i.e., under full solar spectrum irradiation, is calculated as

$$Solar\ AQY\%\ \left(\frac{molecules}{photon}\right) = \frac{k}{A \times \int \Phi d\lambda} \times 100$$

where the photon flux is obtained by applying equation (2) to the standard spectral solar irradiance *H* (ASTM G137-03 AM 1.5G) and integrated over the whole wavelength range (280–2500 nm).

*4.2. Photocatalytic organic reactions.*

Well-defined $Cu_2O$ nanocrystals have been employed in many organocatalytic applications under dark conditions because of the versatile catalytic properties of $Cu^{+1}$ oxidation state [76,77,119]. In addition, $Cu_2O$ nanocrystals have also been applied as photocatalysts for different photocatalytic processes due to the suitable light absorption properties in the visible region of the solar spectrum [15,19,26]. Therefore, a wide variety of photocatalytic reactions have been studied using $Cu_2O$ nanocrystals to understand the effect of different crystal facets. Photocatalytic dye degradation of industrial dye pollutants such as methyl orange (MO) and methylene blue (MB) are used as a common probe reaction to compare the photocatalytic properties of differently exposed crystals facets for $Cu_2O$ nanocrystals [19,39,44,120]. The rate of photocatalytic dye degradation is directly related to the two distinct properties of a semiconductor as follows: 1) strong absorption of the reactant molecules on the catalytic site present in a crystal facet; 2) efficient electron–hole separation upon exposure of light and their fast migration to the active catalytic sites to generate radical species, which then carry out the oxidation of the dye. From a mechanistic point of view, photoexcited electrons reduce the surface adsorbed molecular oxygen ($O_2$) to form the superoxide anion radical ($^{\bullet}O_2^{-}$), which upon protonation generates $^{\bullet}OOH$, $H_2O_2$ and hydroxyl radical ($^{\bullet}OH$) [101]. These radicals may then oxidize the solvated MO molecules (via homogeneous catalysis). The photogenerated holes can either directly oxidize the surface-adsorbed dye molecules (via heterogeneous catalysis) or, more frequently, react with $OH^-$ and $H_2O$ to form $^\bullet OH$ radical, which further carries out the reaction. In addition, the complex interfacial charge transfer between the surface absorbed dye molecules and the photoexcited charge carrier can also lead to an efficient dye degradation [121–123]. However, this kind of mechanism is generally overlooked and needs future investigation.

Huang et al. showed that the rhombic dodecahedral $Cu_2O$ nanocrystal has a higher photocatalytic activity than octahedral, whereas the cubic morphology is completely inactive during the photocatalytic MO degradation [39,44,101]. It is important to note that the amount of catalysts used in the photocatalytic dye degradation process was normalized against a fixed surface area, which is an important parameter allowing to compare the catalytic activity of different well-defined nanostructures. The photocatalytic inactivity of $Cu_2O$ cubic morphology suggested that photogenerated electrons and holes have very strong energy barrier to reach at the active catalytic sites in the exposed {100} surface [39,44,101,120]. Band diagram of cubic $Cu_2O$ nanocrystals shows that the {100} facet has the largest upward band bending (Fig. 18a); for this reason the electron and holes are not able to reach the surface and they undergo recombination, resulting in a negligible photocatalytic activity.

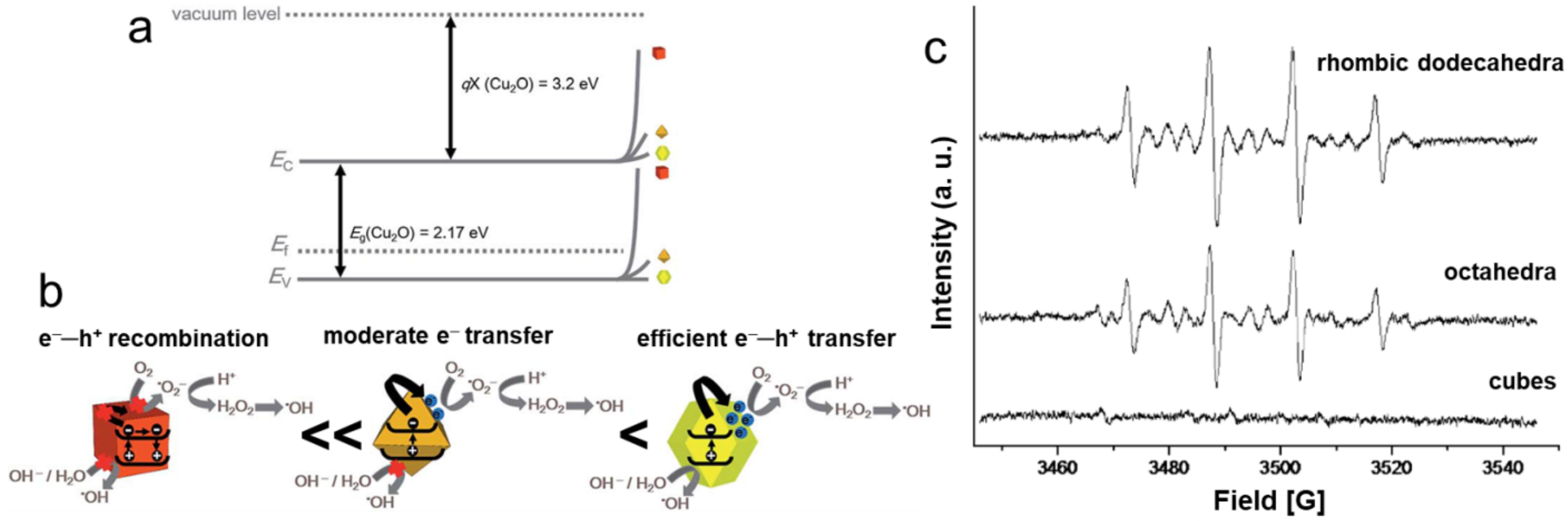

**Figure 18:** (a) Band diagram and corresponding band bending for three shaped $Cu_2O$ nanocrystals. (b) A schematic diagram showing the migration of photoexcited electrons and holes to the three different $Cu_2O$ crystal facets and the corresponding their photocatalytic

activity. Reproduced with permission from ref. [101], Copyright 2017, Royal Society of Chemistry. (c) EPR spectra of DMPO−OH present in photoirradiated $Cu_2O$ cubes, octahedra, and rhombic dodecahedra. Reproduced with permission from ref. [124], Copyright 2016, Wiley-VCH.

To explain the higher photocatalytic activity of rhombic dodecahedral nanocrystals, compared to octahedral, a systematic study has been carried out by adding electron and hole scavengers during the photodegradation of MO [101]. The results indicated that in rhombic dodecahedral nanocrystals bounded with {110} faces, both photoexcited electrons and holes reach the surface active sites more spontaneously and accelerate the formation of a larger number of active $^{\bullet}OH$ radical species, justifying the higher photodegradation rate (Fig. 18b). This {110} facet showed the lowest degree of band bending, which resulted in an improved electron and hole transport to the surface active sites. Octahedral nanocrystals enclosed with the {111} facet had moderate band bending, and the photoexcited electrons were able to reach the surface active sites to produce $^{\bullet}OH$ radical species, as shown in Fig 18b [101]. The photoexcited holes were blocked by the {111} facets of octahedral, and they did not contribute to the production of $^{\bullet}OH$ radicals. The presence of a lower amount of active $^{\bullet}OH$ radical species in the reaction medium was the main reason for the lower photocatalytic activity of octahedral, compared to the rhombic dodecahedral nanocrystals. To further verify this fact, the spin trapping agent 5,5-dimethyl-1-pyrroline N-oxide (DMPO) was introduced to a solution containing three differently shaped $Cu_2O$ nanocrystals, and the electron paramagnetic resonance (EPR) spectra were recorded after the irradiation of light in the presence of oxygen [124]. As shown in Fig. 18c, rhombic dodecahedral nanocrystals produced stronger EPR signals than octahedral nanocrystals, which supported the formation of a higher number of $^{\bullet}OH$ radicals in the presence of rhombic dodecahedral nanocrystals, as discussed above [124]. Most interestingly, photoexcitation of cubic nanocrystals produced no EPR signals, justifying their photocatalytic inactivity. This study strongly demonstrated that the photocatalytic activity of a semiconductor material can be tuned by controlling the exposed facet of a nanocrystal. Most surprisingly, a recent study demonstrated that a dense decoration of 4-ethynylaniline around the cubic $Cu_2O$ nanocrystals altered its band gap and modified the charge carrier transport mechanism through the {100} facet, which led to an enhanced activity [125]. Therefore, the surface-bounded ligands can have a profound impact on the photocatalytic activity of a semiconductor nanocrystal. The photocatalytic dye degradation efficiency of pristine $Cu_2O$ nanocrystals has further been improved by forming heterostructures [126–137]. Among them, $Cu_2O$ nanocrystals decorated with Au, $TiO_2$, and ZnO nanocrystals are the most widely studied solutions. Such hybrid architectures have also been employed to investigate their peroxidase-like activities toward photocatalytic decomposition of different bacteria under visible light illumination [127,138]. All these strategies reduce the photogenerated electron and hole recombination rate and therefore enhance the photocatalytic MO degradation rate by forming a higher number of $^{\bullet}OH$ radicals in the reaction medium. The application of highly oxidizing $^{\bullet}OH$ radicals can be further extended to other industrially important reactions such as photocatalytic alcohol oxidations or alkene epoxidations [139,140]. Thus, well-defined and cheap $Cu_2O$ nanocrystals can show product selectivity in such reactions that need to be studied in the future.

Furthermore, different hybrid nanocatalysts have been developed to provide enhanced activity and stability toward tandem photocatalytic catalytic reactions [141–147]. Knecht et al. deposited Pd nanocrystals on micron sized cubic $Cu_2O$ nanocrystals via a galvanic displacement reaction where surface $Cu^{+1}$ ions became oxidized to $Cu^{+2}$ ions and $Pd^{+2}$ ions were simultaneously reduced to $Pd^0$ nanocrystals [141]. Overnight, the galvanic displacement reaction led to the deformation of the surface {100} planes, followed by the formation of a minor amount of CuO and Cu species at the surface of $Cu_2O$ nanocrystals. The obtained $Cu_2O$–Pd hybrid nanostructures were then employed in a light-driven tandem hydrodechlorination reaction (Fig. 19a). In this process, first, the $Cu_2O$ compartment produced $H_2$ via a photocatalytic water splitting reaction, while in the second step, an in-situ produced $H_2$ was activated by the Pd compartment, where a hydrodechlorination reaction was able to proceed. Conversion of 3-chlorobiphenyl to nontoxic biphenyl in the presence of light is highly important for both organic synthesis and environmental pollution control. No product was obtained when individual $Cu_2O$ nanocubes and Pd nanoparticles were used, suggesting the synergistic effect due to the coupling of these two materials. Decoration of Pd nanocrystals on the cubic $Cu_2O$ nanocrystals reduced the electron–hole recombination as the photoexcited electrons were transferred from the conduction band of the $Cu_2O$ nanocrystals to the Fermi level of the Pd metals. It is interesting to note that the electron transfer become possible through $Cu_2O$ {100} plane to Pd nanocrystals due to the deformation in the exposed crystal plane and change in the surface chemical composition, as stated above.

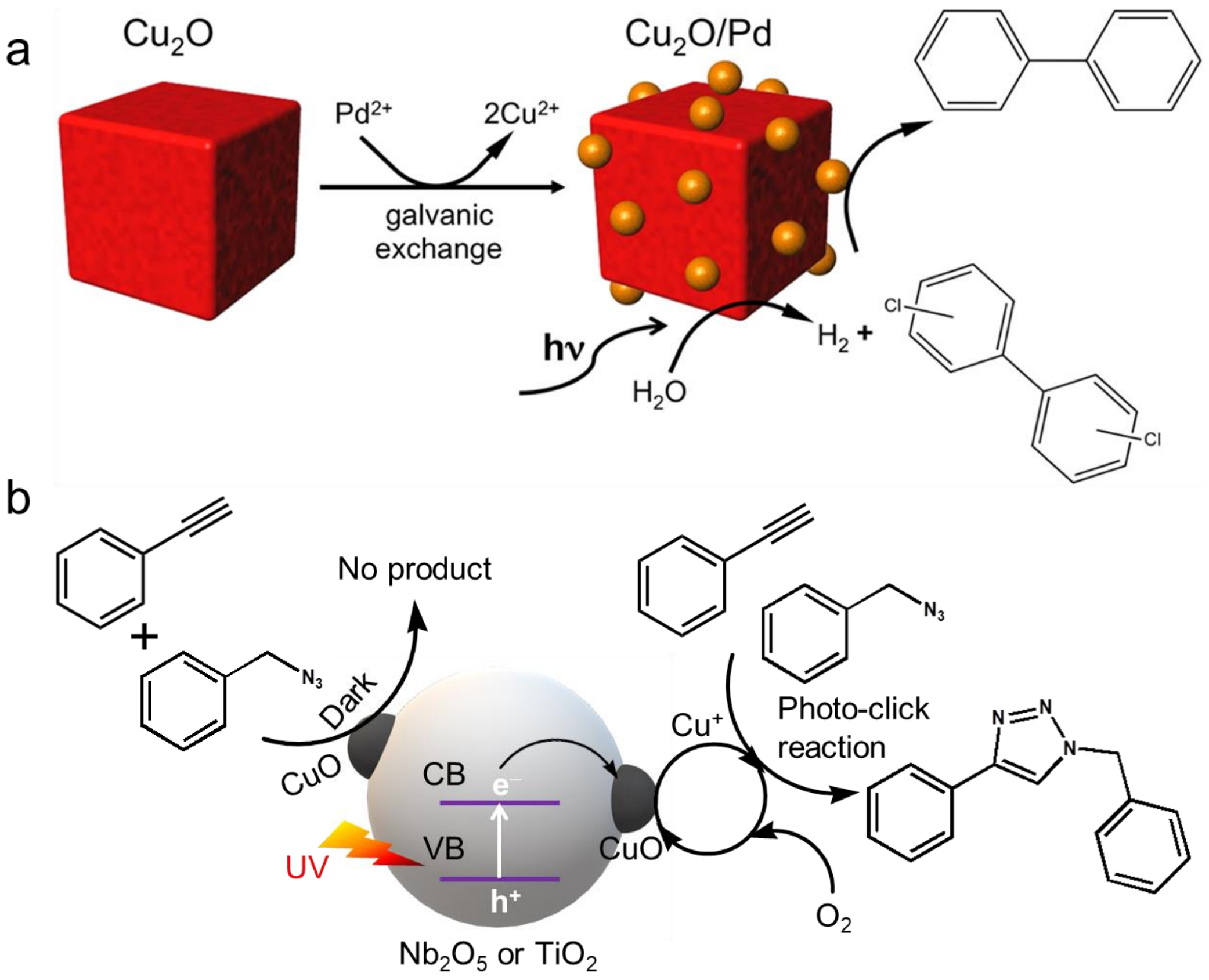

**Figure 19:** (a) Scheme of the synthesis of $Cu_2O$–Pd hybrid photocatalyst for tandem hydrodehalogenation. Reproduced with permission from ref. [141], Copyright 2014, American Chemical Society. (b) Tandem photocatalytic click reaction over semiconductor–CuO nanohybrid. Reproduced with permission from ref. [142], Copyright 2016, American Chemical Society.

Most surprisingly, Scaiano et al. further extended the concept of heterostructure by decorating $CuO_x$ nanoclusters on $TiO_2$ or $Nb_2O_5$ semiconductors whose band gaps fall within the 3−3.5 eV range [142]. These hybrid nanostructures have been used as efficient catalysts for photocatalytic click reactions, especially Huisgen cycloaddition of azides and terminal alkynes to form triazoles, which has many pharmaceutical applications [76]. The click reaction was mainly catalyzed by a $Cu^{+1}$ oxidation state, but it was not carried out by $Cu^{+2}$. Therefore, when UV-light impinged the $TiO_2$ or $Nb_2O_5$ semiconductors, both of them (Fig. 19b) promoted a valence band electron to the conduction band, which was then trapped by surface $CuO_x$ nanoclusters, thus increasing the photoexcited electron–hole separation. After accumulating electrons, $CuO_x$ nanoclusters underwent in-situ reduction, yielding a $Cu^{+1}$ oxidation state, which then efficiently catalysed the click reaction. Once the reaction was completed, the $Cu^{+1}$ oxidation state was regenerated back to a $Cu^{+2}$ state in the presence of oxygen, demonstrating a reversible regeneration of the photocatalyst. Thus, it can be conclude that a suitable support plays an important role in achieving a high photocatalytic activity and stability of the material during the course of reaction [148].

Recently cubic $Cu_2O$ nanocrystals were grown on layered molybdenum disulfide ($MoS_2$) and graphene hybrids, which then were used as photocatalysts for a C–C bond formation reaction by oxidative coupling in the presence of oxygen ($O_2$) [144]. In this case, a question arose: how photoexcited electrons and holes are able to pass via the {100} crystals plane of $Cu_2O$ nanocubes? It is due to the deformation of the {100} crystal plane during the course of the reaction as previously mentioned for Fig. 19a. The $Cu_2O$ nanocubes present in the hybrid nanocomposite have hollow structures and other distorted shapes in minor concentrations. In addition, a longer reaction time, in the presence of $O_2$, can produce the leaching of $Cu^{+1}$ ions from the surface and also considerable changes in the chemical composition and atomic orientation of the {100} crystal plane, which creates this discrepancy. Therefore, researchers must address these crucial points before claiming that $Cu_2O$ nanocubes are highly active photocatalyst for different reactions [149–154]. Recent study showed that when $Cu_2O$ nanocrystals with truncated nanocube morphology were used for the photocatalytic Sonogashira reaction in the presence of $K_2CO_3$ and $CO_2$ atmosphere, a CuO layer was formed in the exposed crystals surface [154]. Another fascinating hybrid photocatalyst where the $Cu_2O$ nanocube was decorated with $SnO_2$ nanoparticles showed a unique post-illumination photocatalytic "memory" effect, which means it can still be catalytically active in the dark after the illumination was switched off for a certain period of time [147]. This "memory effect" was possible due to the presence of reversible chemical states ($Sn^{+2}$ vs $Sn^{+4}$) in $SnO_2$. In this section, we have discussed common $Cu_2O$ photocatalysed organic reactions like dye degradation, click reaction, alcohol oxidation and epoxidation reaction, which are important for environmental applications and industrial organic synthesis, while in the following sections we will turn our attention to photocatalytic reactions that may enable energy transitions such as $H_2$ evolution and $CO_2$ reduction reaction.

*4.3 Photocatalytic $H_2$ evolution reaction.*

Hybrid $Cu_2O$-based nanocrystals are widely investigated for photocatalytic hydrogen evolution [155–177]. In most literature reports, $H_2$ evolution is expressed by $\mu molg_{catalyst}^{-1}h^{-1}$. As we know catalytic reactions take place mainly on the catalyst surface; therefore, it would be more correct to report $H_2$ evolution rates after normalization with surface area i.e. $\mu molm^{-2}h^{-1}$ in addition to $\mu molg_{catalyst}^{-1}h^{-1}$ unit. Such a complete picture will be more useful for comparing $H_2$ evolution rate with other catalysts. However, in reality, there is an almost negligible number of articles that report the $H_2$ production for powder photocatalysts in terms of the surface area. In the future, it is recommended to apply such practice during the demonstration of $H_2$ evolution rate. Similarly, it is also applicable for photocatalytic $CO_2$ reduction reactions, which will be discussed in the following *section 4.4.*

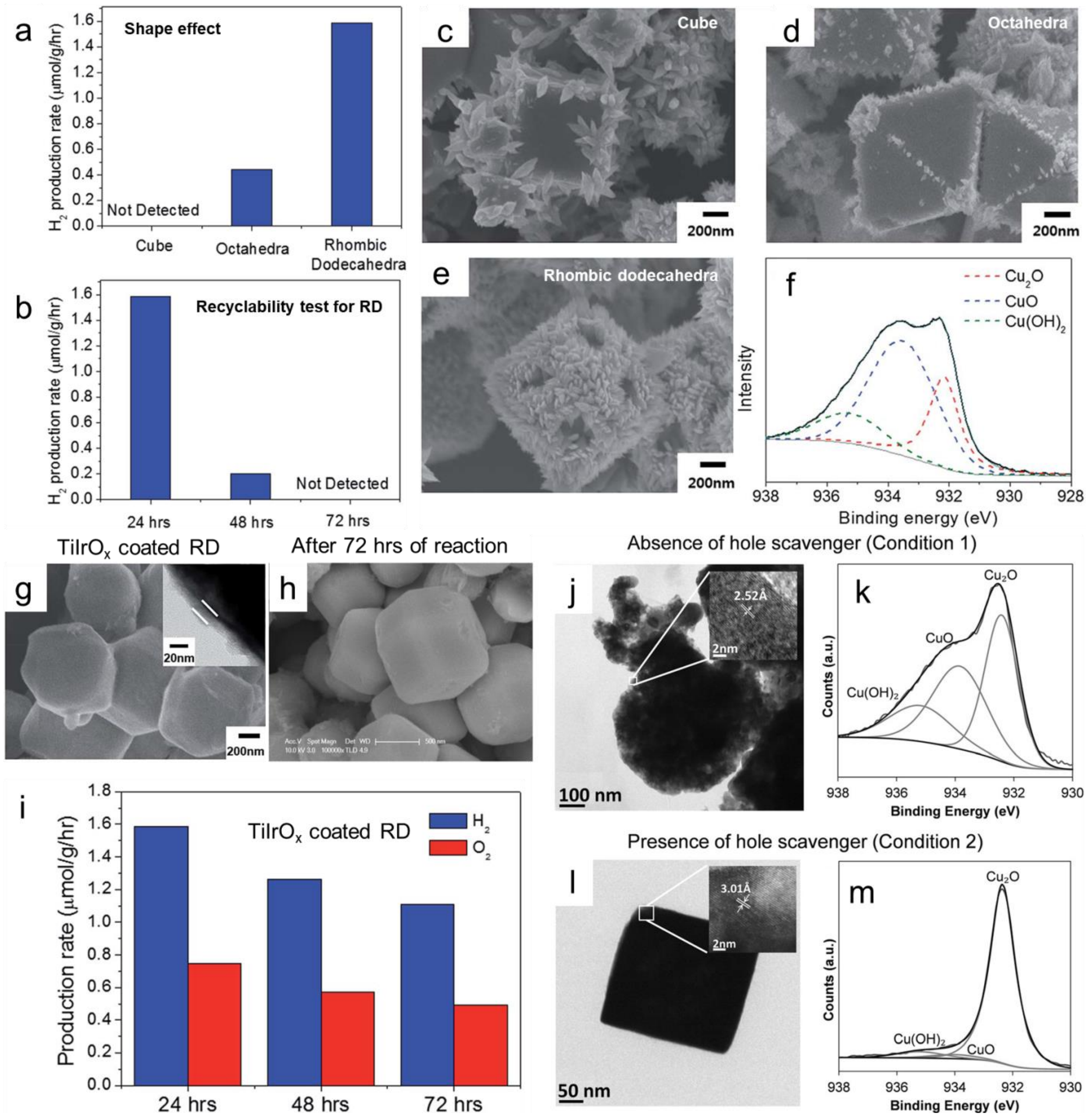

**Figure 20:** (a) Comparison of $H_2$ production rate from overall water splitting using three different $Cu_2O$ morphologies. (b) Recyclable test for $Cu_2O$ RD. (c-e) After 9 h of light irradiation in DI-water without bubbling Ar, SEM images of the three different shapes. (f) Corresponding XPS spectrum of RD. SEM image of $TiIrO_x$ coated $Cu_2O$ RD (g) before, (h) after 72 h of reaction, and (i) their recyclability test. Reproduced with permission from ref. [155], Copyright 2015, Royal Society of Chemistry. (j, l) TEM images and (k, m) XPS of $Cu_2O$ nanocubes under two different conditions. Reproduced with permission from ref. [156], Copyright 2018, Wiley-VCH.

Lee et al. showed that $Cu_2O$ rhombic dodecahedra nanocrystals have higher photocatalytic activity than octahedral shapes whereas cubes are completely inactive for the water splitting reaction (Fig. 20a) [155]. However, RD nanocrystals lose their photocatalytic activity after 24 h of the photocatalytic reaction in pure water (Fig. 20b). A SEM image of these three different morphologies (Fig. 20c–e) proved the different structure distortion as a consequence of the photocatalytic activity.  The RD morphology showed the highest degree of surface distortion and, after use, the highest density of thorns. These thorns are made of CuO as confirmed by XPS (Fig. 20f) and XRD. The presence of any sacrificial reagents is essential to consuming the generated photoexcited holes, with the nature and concentration of these or sacrificial electron donors being relevant for the final performances. In the absence of these sacrificial agents, RD $Cu_2O$ nanocrystals are not able to oxidize water to oxygen; instead, they oxidize $Cu_2O$ to CuO, producing these thorns. While the amount of CuO in the system increases, there is a change in the chemical composition and in the light absorbing properties with a progressive and dramatic decrease in the photocatalytic hydrogen evolution [155]. This observation is very important to understand the photocorrosion properties of $Cu_2O$, which is its weakest point. Increasing the photostability remains a big challenge for the development of low cost efficient photocatalysts for solar to fuel conversion. Different strategies for overcoming the stability issue in $Cu_2O$ systems have been developed, and will be discussed as follows.

*The first strategy* consists in the uniform ultrathin coating of the $Cu_2O$ nanocrystals with a conducting and transparent layer, which allows both electron and hole transfer to its surface active sites, thus minimizing the direct contact between the $Cu_2O$ surface and water while keeping the same photocatalytic activity [155]. Generally, this strategy is more common for $Cu_2O$ based photoelectrode constructions where atomic layer deposition helps to achieve this conformal coating [17,178]. One of such attempts is shown in Fig. 20g, where $Cu_2O$ RD nanocrystals were uniformly coated with a $TiIrO_x$ surface passivation layer ($x < 2$) with a thickness of 20 nm (inset Fig. 20g). Such a layer may consist of a $TiO_x$ layer and $IrO_x$ ultrasmall nanocrystals [155]. The overall water splitting was achieved with a $H_2$:$O_2$ ratio of 2.13 from the $TiIrO_x$ layer coated $Cu_2O$ RD (Fig. 20i) nanocrystals. The $H_2$ production rate was comparable with that of pure rhombic dodecahedral $Cu_2O$, indicating an efficient transfer of electrons to its surface active sites. In the same way, $IrO_x$ facilitated the hole accumulation, thus preventing self-photocorrosion of $Cu_2O$, while enabling water oxidation to produce oxygen. Such techniques also stabilized the shape and morphology of the $Cu_2O$ nanocrystals (Fig. 20h) for a longer time of the photocatalytic overall water splitting reaction with a moderate decrease in the catalytic efficiency (Fig. 20i). Further investigation will be needed to find more efficient materials which will be easy to deposit and can have a profound impact on the photocatalytic efficiency of the $Cu_2O$ nanocrystals with enhanced photostability.

*The second strategy* includes the introduction of suitable hole scavengers in the reaction medium. The hole scavengers are efficiently oxidized by the photogenerated holes in $Cu_2O$ nanocrystals, which in turn prevents the self-photooxidation process. An interesting study showed that when no hole scavengers were added and $Cu_2O$ nanocubes were exposed to pure water in the presence of air and light for a certain period of time, an overall destruction of morphology occurred, as shown in Fig. 20j [156]. The XPS analysis confirmed the $Cu_2O$ self-photooxidation by holes with the formation of CuO and $Cu(OH)_2$ species (Fig. 20k). Most surprisingly, when ethanol was introduced as a hole scavenger into the system, under the same photocatalytic conditions, the morphology and chemical composition remained almost unchanged (Fig. 20 l, m). Due to the favorable redox potential of $Na_2SO_3$, it acts as a more effective hole scavenger than ethanol, resulting in a high photocatalytic hydrogen evolution rate in the presence of $Cu_2O$ nanocrystals with considerable photostability [156].

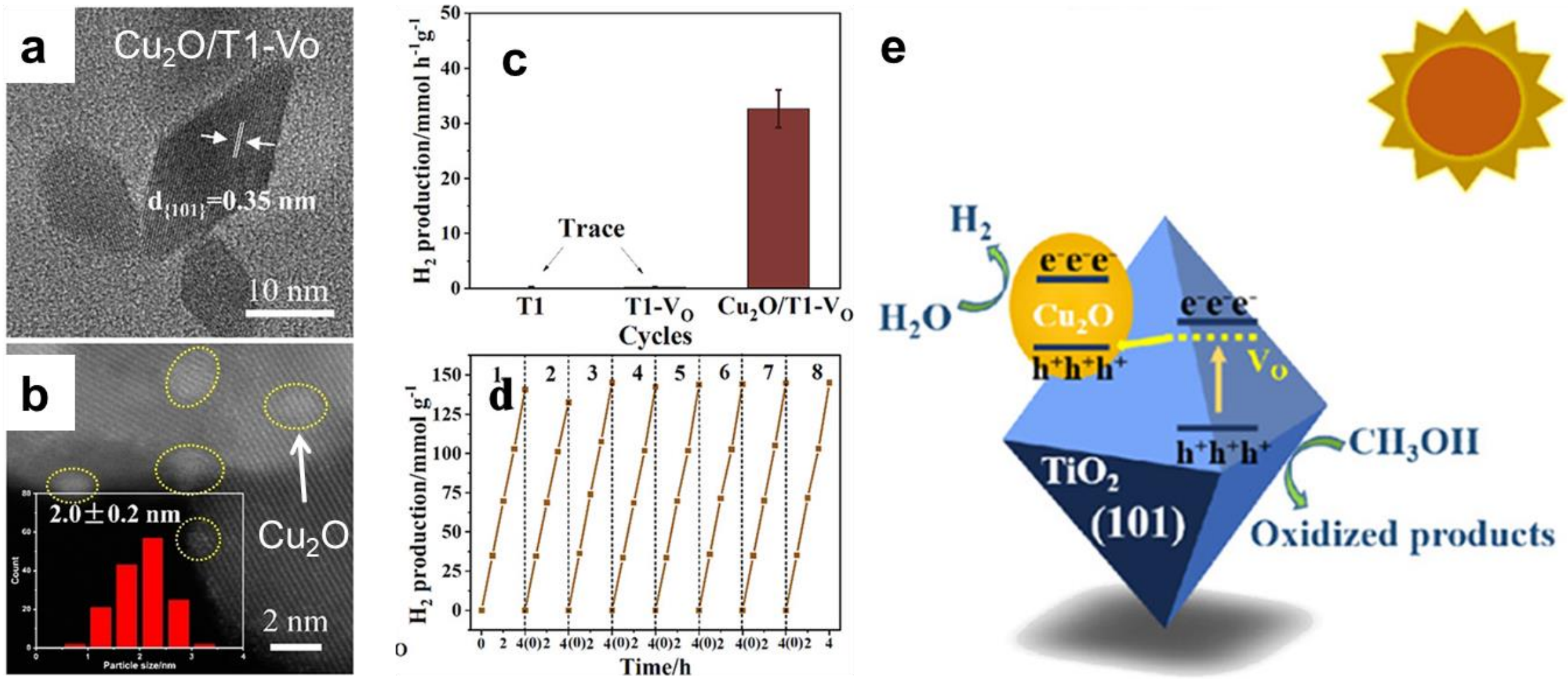

**Figure 21:** (a) HRTEM and (b) HAADF-STEM image of $Cu_2O$ decorated oxygen vacancy ($V_O$) containing 101-faceted $TiO_2$ (referred as $Cu_2O/T1\text{-}V_O$). (c) Comparison of $H_2$ production rates of 101-faceted $TiO_2$ (T1), 101-faceted $TiO_2$ having oxygen vacancies (T1-$V_O$), and $Cu_2O/T1\text{-}V_O$. (d) Recyclability test for $Cu_2O/T1\text{-}V_O$. (e) Graphical presentation for interfacial charge transfer mechanism in $Cu_2O/T1\text{-}V_O$. Reproduced with permission from ref. [163], Copyright 2019, American Chemical Society.

*The third strategy* is the successful establishment of a Z-scheme mechanism that efficiently neutralizes the photogenerated holes in $Cu_2O$, thus preventing its photooxidation [161–166]. Liu group developed a hybrid photocatalyst where 2 nm $Cu_2O$ nanocrystals were uniformly photodeposited on an oxygen vacancy containing 101-faceted octahedral $TiO_2$ nanocrystals (referred as $Cu_2O/T1\text{-}V_O$; Fig. 21a,b) [163]. Under AM 1.5G irradiation, this hybrid catalyst showed 32.6 $mmolg^{-1}h^{-1}$ hydrogen production rate from 10 vol% aqueous methanol solution, which has still been the highest reported rate for $Cu_2O/TiO_2$ hybrid systems to date (Fig. 21c). It also kept its high photostability over the long-term recyclable experiments (Fig. 21d). No oxidized CuO species were detected in the recycled photocatalyst, which confirmed the efficient neutralization of holes in the $Cu_2O$ counterpart via a successful Z-scheme mechanism, as shown in Fig. 21e. An optimum number of surface defects and homogenous distribution of 2 nm $Cu_2O$ nanoclusters, along with 101-faceted $TiO_2$ octahedra, acted synergistically, which resulted in high photoactivity and enhanced photostability of $Cu_2O$ nanoclusters. Based on these findings, further development will be needed to obtain a highly efficient, low cost and stable $Cu_2O$-based photocatalyst for solar $H_2$ production.

*The fourth strategy* is the modification of the $Cu_2O$ nanocrystal surface of carbon-based materials [158,160,167,172,173]. Carbon quantum dots (CQDs) [179], nanodiamond (ND) [158], reduced graphene oxide (rGO) [173], and nitrogen-doped carbon [167] are widely used to enhance the photocatalytic activity and photostability of $Cu_2O$ nanocrystals. Yang et al. showed that the decoration of 50 nm spherical $Cu_2O$ nanocrystals with 3 nm NDs gives a $H_2$ production rate of 1597 $\mu mol g^{-1} h^{-1}$ under simulated solar light irradiation (AM 1.5G, 100 mW $cm^{-2}$, 1 Sun) with a solar-to-hydrogen conversion efficiency of 0.85% [158]. This composite also showed considerable photocatalytic stability. Due to the highly conductive nature of ND and its strong light absorbing properties from UV to NIR regions, the photoexcited electrons from the ND surface were easily injected into the $Cu_2O$, and, simultaneously, the photogenerated holes from $Cu_2O$ were being injected to ND. The excess holes on the ND surface carried out the photooxidation of ethanol, whereas the excess electrons in $Cu_2O$ carried out the photoreduction of $H^+$ to $H_2$. Such spontaneous separation and consumption of photoexcited electrons and holes were the main reasons for the enhanced photostability of ND decorated $Cu_2O$ hybrid systems. In another study, glucose was used as the carbon precursor to form a protective 20 nm thick carbon layer coated on a $Cu_2O$ nanowire, which resulted in enhanced photostability and a water splitting performance [172]. The photoexcited electrons were easily transported to the reaction medium via this carbon layer, whereas, because of its hydrophobic nature, it protected $Cu_2O$ from $H_2O$.

However, organic molecules such as ethanol or methanol make their passage to the $Cu_2O$ surface through this carbon layer where they become easily oxidized by photogenerated holes, resulting in enhanced photostability. Such an approach is highly cost-effective and easily scalable; the only thing that needs to be taken into account is how to achieve the optimum thickness of the carbon layer on the $Cu_2O$ nanocrystals.

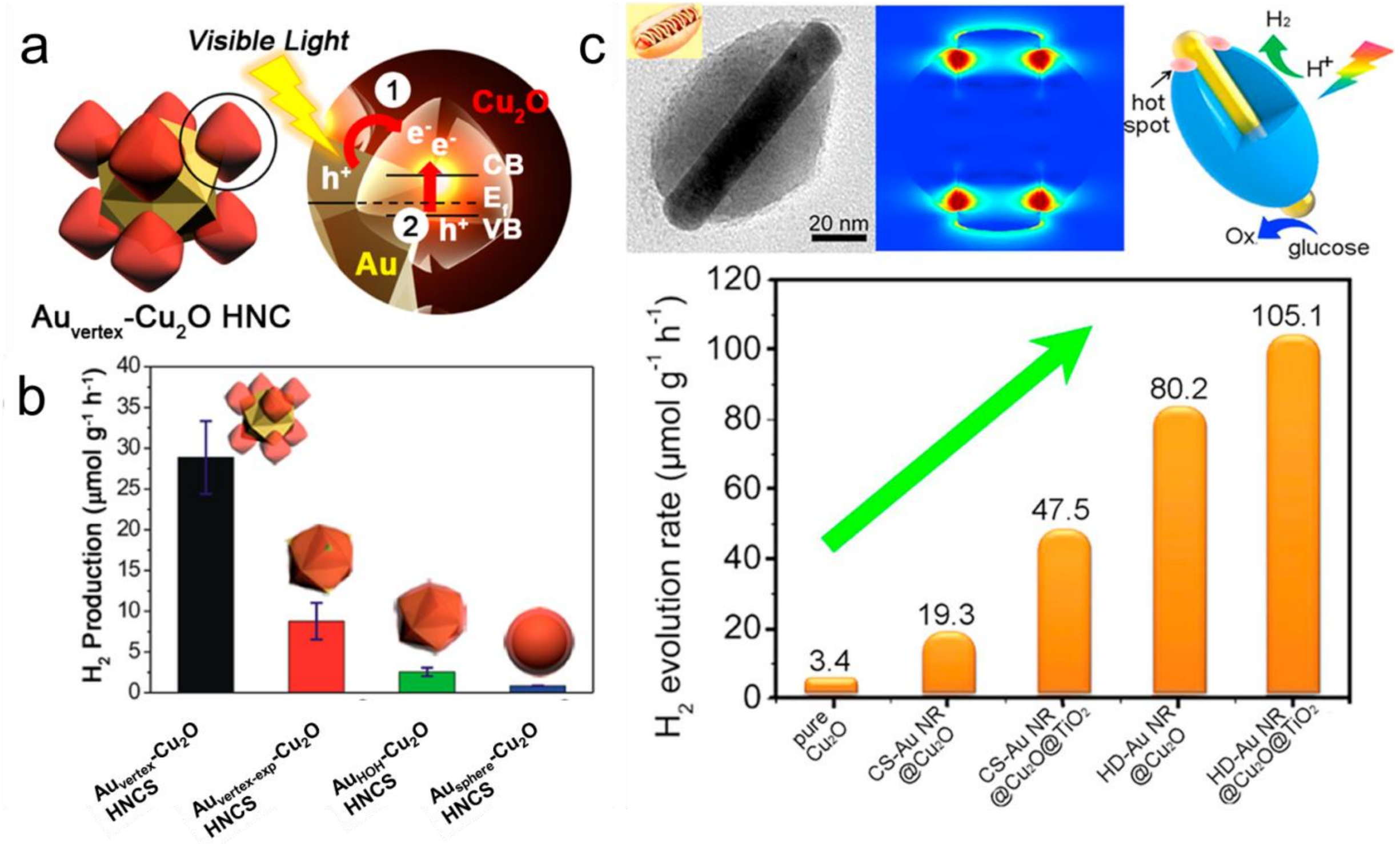

**Figure 22:** (a) Schematic model for $Au_{vertex}$-$Cu_2O$ heteronanocrystals (HNC) and its SPR-induced hot electron transfer mechanism. (b) Comparison of photocatalytic $H_2$ evolution over different HNCs. Reproduced with permission from ref. [169], Copyright 2016, American Chemical Society. (**c**) An overall schematic diagram of hot-dog like Au nanorod@$Cu_2O$@$TiO_2$ heteronanostructure with comparison to hydrogen production activity. Reproduced with permission from ref. [170], Copyright 2017, Elsevier.

*A fifth strategy* foresees the formation of heterostructures with plasmonic nanocrystals with sharp vertices and edges [168–171]. In this regard, Han group used hexoctahedral (HOH) Au nanocrystals enclosed by 48 triangular high-index {321} facets as plasmonic antenna and selective overgrowth of $Cu_2O$ (reactive site) was achieved on the vertices of the HOH Au NCs with the help of facet-selective adsorption behavior of PVP surfactant (Fig. 22a) [169]. Such hetero-nanocrystals showed a higher photocatalytic $H_2$ production rate than core-shell nanostructures (Fig. 22b). Such an enhanced rate for $Au_{vertex}$-$Cu_2O$ HNCs could be attributed to the LSPR-induced hot electron transfer from HOH Au nanocrystals to $Cu_2O$. An FDTD simulation confirmed that strong electromagnetic field enhancements were mainly localized at the vertices of HNCs where $Cu_2O$ nanoclusters were mainly present, and an efficient hot electron transfer took place, facilitating the proton reduction process. Due to the formation of a suitable Schottky junction between the Au and $Cu_2O$, the photogenerated holes from the $Cu_2O$ nanoclusters efficiently migrated to Au nanocrystals where methanol scavenged the holes spontaneously due to the availability of the exposed Au surfaces in $Au_{vertex}$-$Cu_2O$ HNCs [169]. These kinds of HNCs also contribute to increasing the photostability of $Cu_2O$. Different exotic structures such as hot-dog like Au nanorod@$Cu_2O$@$TiO_2$ (Fig 22c) nanocrystals and other hybrid systems showed enhanced photocatalytic activity with increased photostability [170,171,180–182].

*Finally, the sixth strategy* is based on the careful selection of reagents used during the photocatalytic process. As we have discussed earlier in *Section 2.3*, the presence of HCl; $NH_3$; or other strong bases such as $S^{-2}$ or $S_2O_3^{-2}$ anions, and dissolved $O_2$, can seriously cause distortion or etching of the $Cu_2O$ nanocrystals. Therefore, they should be avoided along with the use of short reaction times and low temperatures in order to increase the photostability of $Cu_2O$ nanocrystals. Dang et al. recently showed that in the presence of $Na_2SO_3$ as a hole scavenger, $Cu_2O$@$Cu_7S_4$ core@shell nanocubes showed an efficient photocatalytic hydrogen production with enhanced photostability [171]. To sum up, studies of different hybrid photocatalysts and their electron transfer mechanisms provide a fundamental understanding of how to enhance the photocatalytic efficiency and photostability of $Cu_2O$ by suppressing its self-photooxidation under light illumination.

*4.4 Photocatalytic $CO_2$ reduction reaction.*

The photocatalytic reduction of $CO_2$ with $H_2O$ to solar fuels is considered an artificial photosynthesis reaction. It attracted huge attention due to the recent global warming problems, and motivated the attempt to reduce our dependence on fossil fuels [183–186]. Using water as the source of $H_2$ instead of pure $H_2$ gas during this photoreduction process is the most desirable and challenging task. If we can develop such an energy-efficient system, it may be the most sustainable and environmentally green process to complete the carbon cycle [187–191]. Major difficulties associated with this photoreduction process are the complex reaction mechanism, which involves different modes of adsorption of $CO_2$ molecules on the catalyst surface and the multi electron-multi protonation coupling-decoupling steps with high activation energy barriers for different intermediates [183]. These reaction steps make this reaction photocatalytically less efficient and also lead to generation of a wide variety of products such as CO, $CH_3OH$, HCHO, HCOOH, and $CH_4$ (generally termed as C1 products). Complex in-situ C−C coupling and further modification of C1 products generates compounds with two carbons (C2 products) such as $C_2H_6$, $C_2H_4$, $CH_3CH_2OH$, $CH_3CHO$, and $CH_3COOH$ [183–186]. Due to the high activation energy barrier, C2 products are more difficult to form, as compared to C1 products, during the photocatalytic reduction reaction. However, C2 products are of higher importance in the fields of chemical, fuel and polymer industries. Further coupling of C1 and C2 products can lead to the formation of C3 products, which is rarely formed via a photocatalysed reaction over Cu–based catalysts, but it can be achieved via an electroreduction reaction [192,193]. Therefore, better understanding of the reaction mechanism is necessary in order to improve the product selectivity along with achieving high conversion rates. In this regards, Mul et al. showed that the photocatalyst prepared with high molecular weight carbon containing precursors or solvents were contaminated with small quantities of carbon residues even after high temperature calcination as the final stage [12]. These carbon impurities have a profound impact on the photocatalytic water activation and $CO_2$ reduction reaction. Figure 23a shows in-situ DRIFT spectra of the cuprous oxide/$TiO_2$ catalyst after 80 min of photoirradiation under different gaseous atmospheres. When isotopically labeled $^{13}CO_2$ and $H_2O$ were used, the ratio $^{12}CO$: $^{13}CO$ was surprisingly 6:1, which indicated that the carbon residues on the catalyst surface were involved in the reactions with $^{13}CO$ and $H_2O$ molecules in the presence of light, and produced $^{12}CO$ in excess. This is not coming from artificial photosynthesis reaction, as shown in the following equation:

$$^{13}CO_2 + {}^{12}C \rightarrow {}^{13}CO + {}^{12}CO \quad (1)$$

$$H_2O + {}^{12}C \rightarrow {}^{12}CO + H_2 \quad (2)$$

Many recent studies report a high activity with product selectivity for different photocatalysts. Nevertheless, in the future, authors should pay extra attention, before publishing their data, to ruling out any overestimation of the reaction rate coming from carbon contamination present in the photocatalyst surface. This overestimation can be caught by simply carrying out a blank photocatalytic test in the absence of $CO_2$ but in the presence of $H_2O$ [12]. In addition, photoreduction should also need to be carried out using isotope labeled $^{13}CO_2$ for further confirmation of the product selectivity and reaction rate. And lastly, it is also necessary to follow the previously described surfactant removal steps (*Section 2.3*) during the synthesis of photocatalyst in order to get rid of any unwanted carbon contamination.

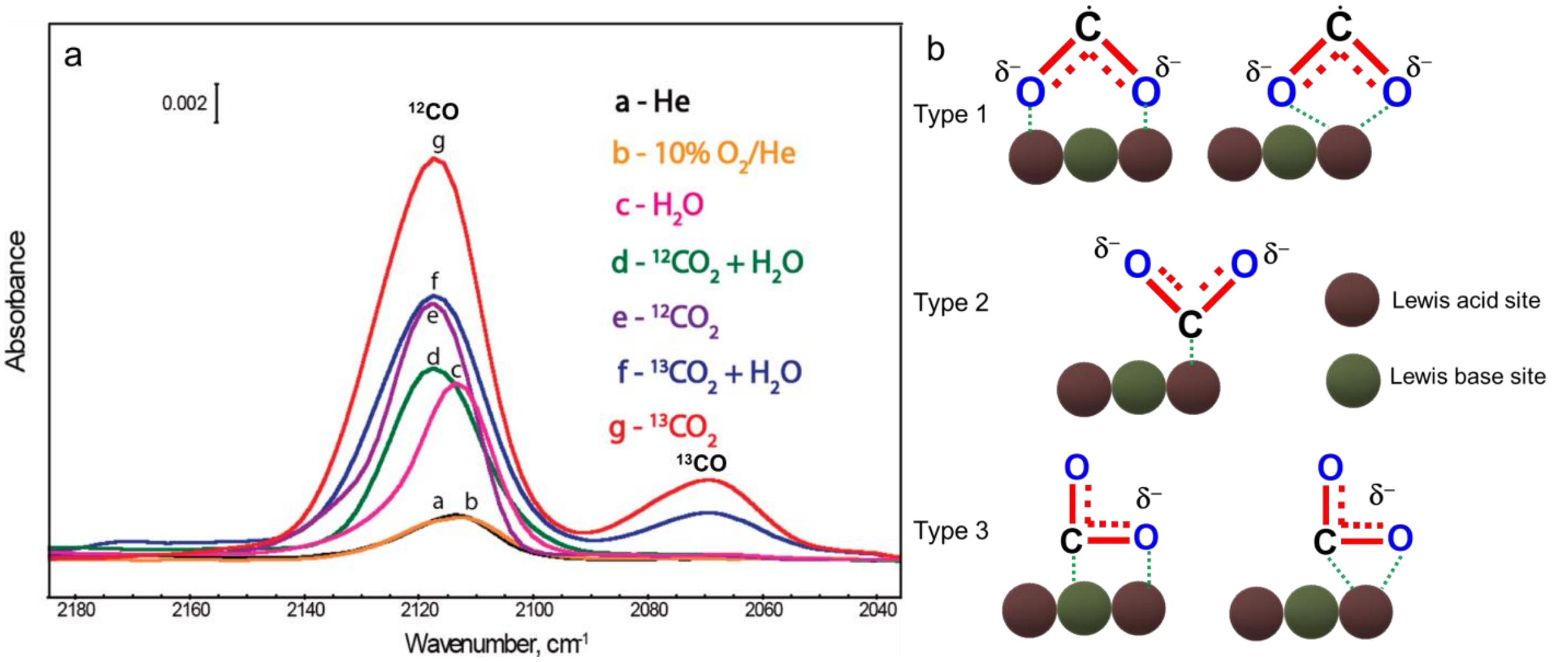

**Figure 23:** (a) Role of carbon residues on catalyst: DRIFT spectra showing the distribution of products under different photoreduction conditions. Reproduced with permission from ref. [12], Copyright 2010, American Chemical Society. (b) Different binding mode of $CO_2^{\bullet-}$ radical on the catalytic site. Reproduced with permission from ref. [183], Copyright 2019, American Chemical Society.

The photoreduction of $CO_2$ with $H_2O$ to these products is a thermodynamically uphill conversion with a positive Gibbs free energy, ΔG, which makes it difficult to proceed at room temperature and without any supply of high external energies. Therefore, it is of great importance to develop efficient semiconductor-based photocatalysts to achieve a desirable photocatalytic conversion rate with high product selectivity. Different ways have been considered to modify these semiconductors by the fine tuning of the band gap, morphology, and size, and with addition of suitable co-catalysts [183–191]. Under suitable light illumination, photoexcited electrons present on the reduction site in a hybrid catalyst typically reduce $CO_2$ to solar fuels, whereas the photoexcited holes presents on oxidation site react with water to produce $O_2$. For an optimum performance, the CB position of the hybrid catalyst must be much higher than the proton-assisted multielectron reduction potentials of $CO_2$ to different products, whereas the VB edge must be much more positive than the water oxidation potential (Fig. 2) [183–186]. Apart from an optimum control of the semiconductor band gap and energy alignment, a right choice of the co-catalyst is also necessary for enhancing the adsorption and activation of $CO_2$ molecules on the active sites, which in turn accelerates the photoreduction of $CO_2$ with high product selectivity [183]. As shown in Fig. 2, the photoreduction of $CO_2$ to $CO_2^{\bullet-}$ radical has large reduction potential (−1.49 V versus NHE, pH=0) and also its geometry changes from linear to a bent shape; an increase in steric hindrance will make this preliminary one electron reduction step highly unfavorable. The formation of destabilized bent $CO_2^{\bullet-}$ radical is the most important step for initiating multistep photoreduction of $CO_2$ and it can be stabilized on the surface active sites via three different coordination modes, as shown in Fig. 23b [183]. Different chemical composition, morphology, electronic band structure, and the exposed crystal planes have a significant impact on these different kinds of coordination modes, which in turn have a direct effect on the $CO_2$ photoreduction rate and product selectivity. Extensive theoretical work by Carter group shows that $Cu_2O$ is a promising photocatalyst for photoconversion of $CO_2$ to methanol [194–196]. Leeuw et al. performed an in-depth DFT calculation for different binding modes of $CO_2$ molecules on three low-index $Cu_2O$ crystal planes and their corresponding adsorption energy [197]. As shown in Figure 24, the $Cu_2O$ (110):Cu–O crystal plane shows the strongest chemisorption of the $CO_2$ molecule, releasing ∼170 kJ/mol energy, which supports the formation of bent $CO_2^{\bullet-}$ radical most efficiently. On the other hand, $Cu_2O$ (111):O surface shows the weakest adsorption, and, therefore, the $CO_2$ molecule remains linear [197]. The activation of the $CO_2$ molecule is directly proportional to the adsorption energy. The higher the adsorption energy of a $CO_2$ molecule on a particular facet of the $Cu_2O$ nanocrystal, the easier will be the formation of the bent $CO_2^{\bullet-}$ radical anion. These different adsorption energies and their conformation can directly influence the reaction rate and product selectivity.

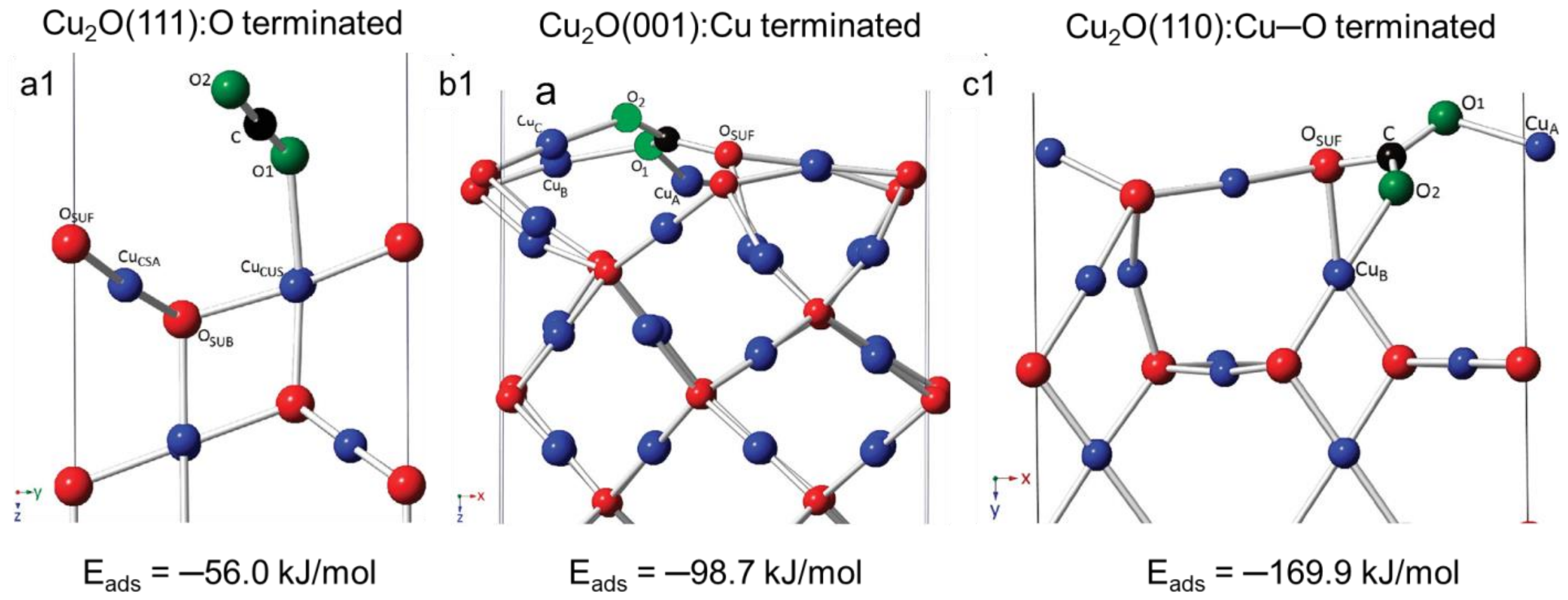

**Figure 24:** Crystal structures of $Cu_2O$ for three different surface planes (a1) {111}, (b1) {001} and (c1) {110} in their most stable configuration respectively. Blue and red spheres are copper and oxygen atoms respectively. (a2–c2) Different binding mode of $CO_2$ molecules on these crystals planes with their adsorption energy respectively. Black and green sphere represent carbon and oxygen atom in a $CO_2$ molecule. Reproduced with permission from ref. [197], Copyright 2016, American Institute of Physics.

Cubic, octahedral, and rhombic dodecahedral $Cu_2O$ crystals decorated with reduced graphene oxide (rGO) nanosheets were recently synthesized, and their photoreduction of $CO_2$ in aqueous solution was compared [198]. It is interesting to notice that all of these hybrid photocatalysts exclusively produced $CH_3OH$, and other possible products such as HCHO and HCOOH were not detected in the system. The methanol production rate followed the order: cube-$Cu_2O$/rGO < octahedral-$Cu_2O$/rGO < rhombic dodecahedral-$Cu_2O$/rGO. Considering Figure 18 and 24, the highest photo-conversion of $CO_2$ to MeOH for RD could be explained by two points: 1) easy accumulation of photoexcited electrons on the {110} surface and 2) easy formation of the bent $CO_2^{\bullet-}$ radical anion. Wrapping with rGO provided higher stability for the $Cu_2O$ nanocrystals during the course of the reaction and also better electron-hole separation. Recently, Rajh et al. used in-depth characterization techniques such as correlated scanning fluorescence X-ray microscopy, EPR, and environmental TEM at atmospheric pressure to prove that the (110) facet of $Cu_2O$ was more active in the photoreduction of $CO_2$ with $H_2O$ to methanol, while the (100) facet was inert during this process [199]. Methanol was produced exclusively as the main product, which could be correlated to the fast desorption of methanol from the (110) facet. No other oxidized products such as HCHO and HCOOH were found in the system. This further emphasizes that the facet-dependent photoreduction of $CO_2$ requires a precise control of the shape and size of the nanocrystal to achieve a high photoreduction rate and product selectivity, even more than in the case of $H_2$ production.

In addition to pristine $Cu_2O$ nanocrystals, different hybrid system have also been developed in order to enhance their better charge separation and product selectivity during the photoreduction of $CO_2$ [179,200–211]. A novel carbon quantum dots (CQDs) decorated $Cu_2O$ heterostructures were synthesized via a one-pot procedure, where ~ 5 nm CQDs were homogeneously decorated on the surface of a ~2 μm spherical crystalline $Cu_2O$ particle [179]. This hybrid nanostructure showed strong absorption in the complete solar spectrum range with a band gap 1.96 eV, much lower than that of pure bulk $Cu_2O$ (2.2 eV). The photoexcited electrons in the $Cu_2O$ microsphere exclusively reduced $CO_2$ into $CH_3OH$, whereas the holes were transferred to CQDs and oxidized $H_2O$ to $O_2$. Thus, the CQDs co-catalyst enhanced both the photocatalytic performance and the stability of $Cu_2O$ microspheres. Recently, 4.4 nm $Cu_2O$ nanocrystals supported on defective graphene were used as an efficient photocatalyst for gas-phase methanation of $CO_2$ in the presence of $H_2$ at high temperature, also known as the Sabatier reaction [200]. This hybrid photocatalyst showed exclusive $CH_4$ production with an AQY at 250 °C of 7.84% in the 250–360 nm wavelength range. In the third cycle of the reaction, the photocatalytic activity decreased significantly because of the complete conversion of $Cu_2O$–graphene to Cu–graphene. This conversion was not probably due to the photocatalytic process itself but it was related to the high temperature reductive environment. Doping also has been used as an important tool to tune the band gap of a semiconductor and to enhance the photocatalytic performance and product selectivity. For instance, Cl-doping of $Cu_2O$ nanorods has been used to shift the band structure of $Cu_2O$ to a more positive valence-band position, which facilitates the water oxidation [201]. The Cl-doped $Cu_2O$ nanorods showed excellent photoreduction of $CO_2$ in the presence of $H_2O$ under visible-light irradiation with selective formation of CO.

Another possible way to increase the photoreduction performance of cuprous oxide is the suitable formation of Z scheme-based hybrids by coupling p-type $Cu_2O$ semiconductor with n-type wide bandgap semiconductor such as $TiO_2$, ZnO and $W_{18}O_{49}$ [203,204,212–215]. Precise control of the morphology and size can efficiently increase the photoreduction rate and product selectivity. A Z-scheme heterojunction can limit the issue of recombination of the charge carriers, improving the stability of the photocatalyst.

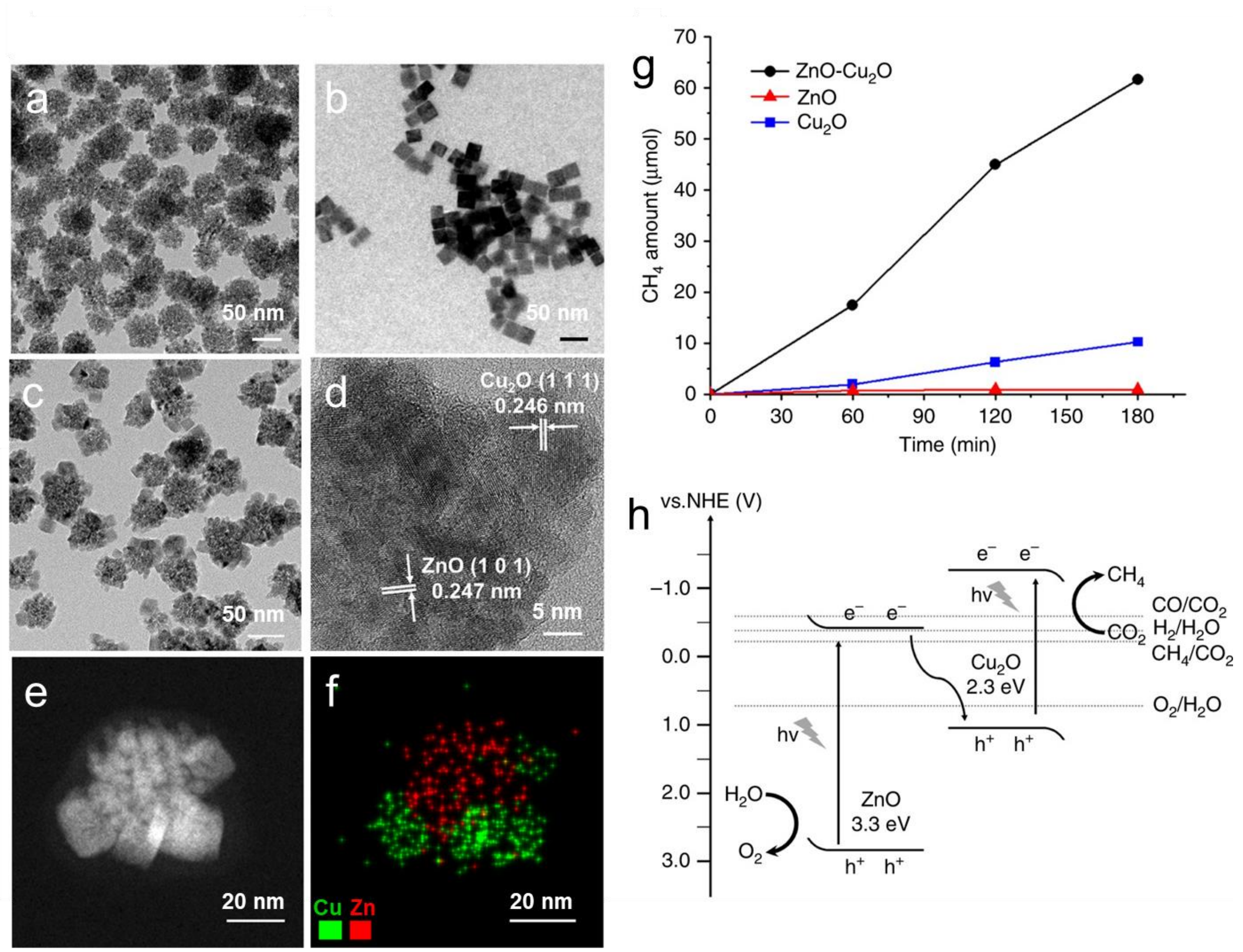

**Figure 25:** TEM images of (a) ZnO spheres, (b) $Cu_2O$ nanocubes, (c) ZnO-$Cu_2O$ hybrid nanoparticles, (d) HRTEM, (e) STEM image, and (f) elemental mapping of an individual ZnO-$Cu_2O$ nanoparticle. (g) Comparison of photoreduction of $CO_2$ using this three different catalyst. (h) Band alignment and proposed electron transfer mechanism of the ZnO-$Cu_2O$ hybrid catalysts. Reproduced with permission from ref. [204], Copyright 2017, Springer Nature.

Grela et al. used octahedral $Cu_2O$ microcrystals and decorated its 8 {111} crystal planes with $TiO_2$ nanocrystals uniformly [203]. When this octahedral $Cu_2O$–$TiO_2$ hybrid photocatalyst was irradiated with UV-Visible light ($\lambda \geq 305$ nm) in the presence of water vapor and $CO_2$, the photoreduction reaction proceeded with selective formation of CO. The role of carbon contamination in the catalyst surface was not properly investigated in this study. In addition to this, the octahedral $Cu_2O$–$TiO_2$ hybrid system also showed enhanced photo-stability in comparison with pure $Cu_2O$, which was due to the efficient transfer of photoexcited CB electrons to the photoexcited VB holes of $Cu_2O$. Thus, $TiO_2$ also protected $Cu_2O$ from photocorrosion during the course of reaction. These results provided a direct evidence of an efficient Z-scheme, which was further confirmed by an in-situ analysis of the photocatalyst using XPS and EPR techniques. In addition, Song et al. also developed a hybrid photocatalyst where 40 nm ZnO nanosphere was decorated with multiple 18 nm cubic $Cu_2O$ nanocrystals, as shown in Figure 25c [204]. These ZnO-$Cu_2O$ hybrid photocatalysts showed a selective $CH_4$ formation with estimated quantum efficiency (QE) of 1.5% using 200 to 540 nm wavelength region. A synergistic charge transfer between the ZnO and $Cu_2O$ counterparts led to an enhanced $CO_2$ photoreduction rate (Fig. 25g). To understand the enhanced photocatalytic activity, an identical reaction performed under visible light (UV cutoff filter $\lambda > 420$ nm) was performed, and a negligible amount of $CH_4$ was obtained. When the cutoff filter was removed, the formation rate for $CH_4$ was similar to that observed for the original reaction, which proved that the operational Z-scheme mechanism under UV-Visible light illumination (as shown in Fig 25h) had caused the hybrid photocatalyst to outperform their individual component. More diverse combination of hybrid catalysts with different shapes and sizes need to be developed so that they can operate via the Z-scheme mechanism, thus improving the photostability of the $Cu_2O$ counterpart with enhanced photoreduction rate and product selectivity. Chen et al. used a simple self-assembly strategy to decorate $Ti_3C_2$ QDs onto $Cu_2O$ nanowires to a achieve selective formation of methanol [205]. The photoexcited CB electrons of $Cu_2O$ were readily transferred to the attached $Ti_3C_2$ QDs, as the latter is a highly conductive material having the $E_F$ less negative than the CB of $Cu_2O$, which enhances the charge separation. Similarly, as the $E_F$ of $Ti_3C_2$ QD is more negative than the redox potential of $CO_2$ to methanol, the accumulated electrons efficiently performed the photoreduction reaction.

Apart from the Z-scheme mechanism, the construction of hybrid 'antenna-reactor' photocatalysts also attracted huge attraction due to their enhanced photocatalytic performance compared to their individual components [118,206,207,216]. In this system, generally plasmonic materials (such as Au, Ag, Cu, TiN, and Al) act as an 'antenna', which mainly absorbs the light and generates hot electrons (and eventually heat). These hot electrons are then transferred to the 'reactor' site (usually formed by transition metals such as Pt, Pd, Rh, Ru), which is mostly responsible for performing the photocatalytic reaction.

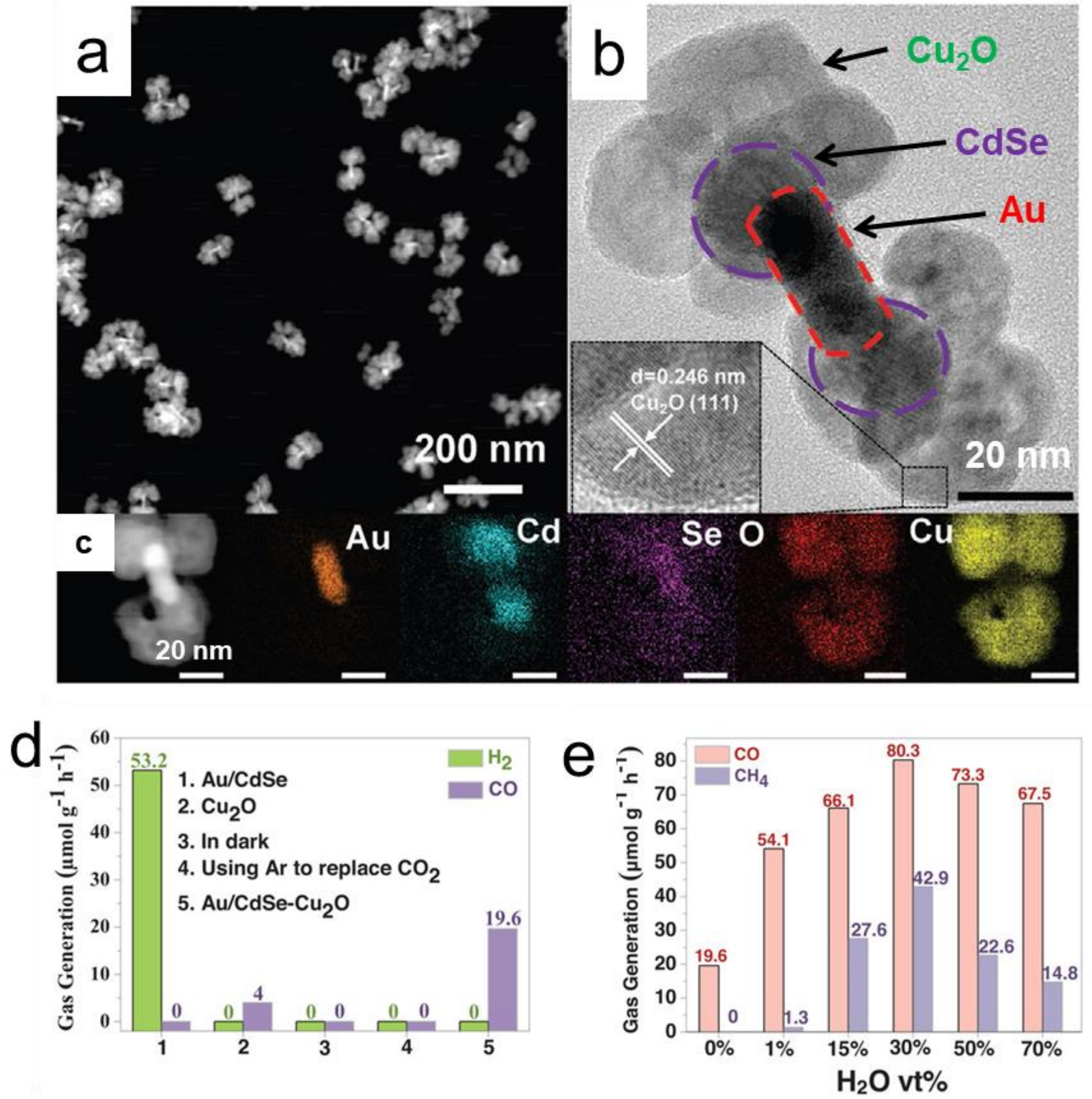

**Figure 26:** Au/CdSe–$Cu_2O$ hybrid nanostructures (a) STEM image, (b) HRTEM image, (c) STEM image of one nanostructure, and its corresponding EDS elemental mapping. Control photocatalytic $CO_2$RR experiment for (d) different samples without water (e) with different volume of water. Reproduced with permission from ref. [207], Copyright 2020, Wiley-VCH.

Halas et al. developed a Al–$Cu_2O$ core–shell hybrid photocatalyst where Al was used as a low cost and earth abundant plasmonic core, while the polycrystalline $Cu_2O$ shell acted as a catalyst for the reverse water-gas shift (rWGS) reaction ($CO_2+H_2 \leftrightarrow CO+H_2O$) [206]. Interestingly, at higher illumination intensities, the Al–$Cu_2O$ core–shell hybrid photocatalyst showed a much higher CO formation rate than the individual $Cu_2O$ and the Al counterparts. Most surprisingly, under a light driven rWGS reaction, a highly selective formation of CO took place, whereas a purely thermal-driven reaction lost the product selectivity as both $CH_4$ and CO were found in the product mixture. Zhang et al. used a three hetero-compartment hybrid photocatalytic system where the $Cu_2O$ hemisphere selectively covered the CdSe part of a plasmonic Au/CdSe nanodumbbell, as shown in Fig. 26a–c [207]. It is interesting to note that the use of PEG, methyl ether, and hydrazine as reducing agents led to the formation of this exotic kinds of hybrid nanostructures. Figure 26d demonstrates that when the photoreduction of $CO_2$ was carried out in a mixed organic solvent ($CH_3CN$:TEOA = 9:1) without adding any water under visible light illumination, the Au/CdSe nanodumbbell generated only $H_2$ without any detectable CO or $CH_4$. The protons for hydrogen production might come from a trace of water in $CH_3CN$. However, when $Cu_2O$ fully covered the CdSe hemisphere, it was found a 100% CO selectivity without any $H_2$. The addition of different amounts of $H_2O$ to the catalytic system systematically increased the production of both CO and $CH_4$ (Fig. 26d). The addition of $H_2O$ to the mixture of the $CH_3CN$ and TEOA solvents catalyzed the formation of $CH_4$ (a $8e^-/8H^+$ pathway) and CO (a $2e^-/2H^+$ pathway), by providing more protons and electrons into the reaction mixture. The exact electron transfer mechanism for this enhanced photocatalytic preformation is still under investigation.

As discussed above, hydrogen evolution appears as a competitive reduction reaction during the photoreduction of $CO_2$. Thus, is it necessary to suppress the $H_2$ evolution so that we can use the photoexcited electrons more efficiently for the generation of other liquid or gaseous carbon-based solar fuels. The most studied and commercially important $TiO_2$ (Degussa P25), loaded with Pt, leads to hydrogen evolution during the photoreduction of $CO_2$ in the presence of water, with small fractions of CO and $CH_4$ [208]. Interestingly, Wang and his coworkers found that when a $Cu_2O$ shell was selectively photodeposited on the Pt nanocrystals, it enhanced the production of $CH_4$ and CO over $H_2$ (Fig. 27a). By systematically increasing the Cu content in the catalyst, the selectivity for $CO_2$ reduction increased from 30% to 85% (Fig. 27b). As no high-molecular-weight organic surfactant or solvent was used during the photodeposition process, no carbon contamination was detected on the surface of the final catalyst, and the product selectivity purely came from the photocatalytic conversion of $CO_2$ [12].

Inspired by this work, a Yan's group demonstrated a facet-dependent Schottky heterojunction by coupling $BiVO_4$ truncated octahedron with a hemispherical Au@$Cu_2O$ core-shell cocatalyst [210]. Truncated octahedron consist of anisotropic facets such as {010} and {110}. Photoexcited electrons can be selectively accumulated on the {010} facet, whereas holes become accumulated on the {110} facet. Therefore, Au nanocrystal can be selectively deposited on the {010} facet by photoreduction. By contrast, before Au nanocrystals deposition only on the {110} facet, a PVP surfactant is needed to block the {010} facet by preferential adsorption. The $Cu_2O$ shell layer was subsequently deposited on Au via a simple chemical reduction, forming three different heterostructures such as $BiVO_4$\{010\}-Au-$Cu_2O$, $BiVO_4$\{110\}-Au-$Cu_2O$ and $BiVO_4$\{010\}-$Cu_2O$ (as shown in Fig 27c). Such synthetic strategies are very challenging and need special attention.

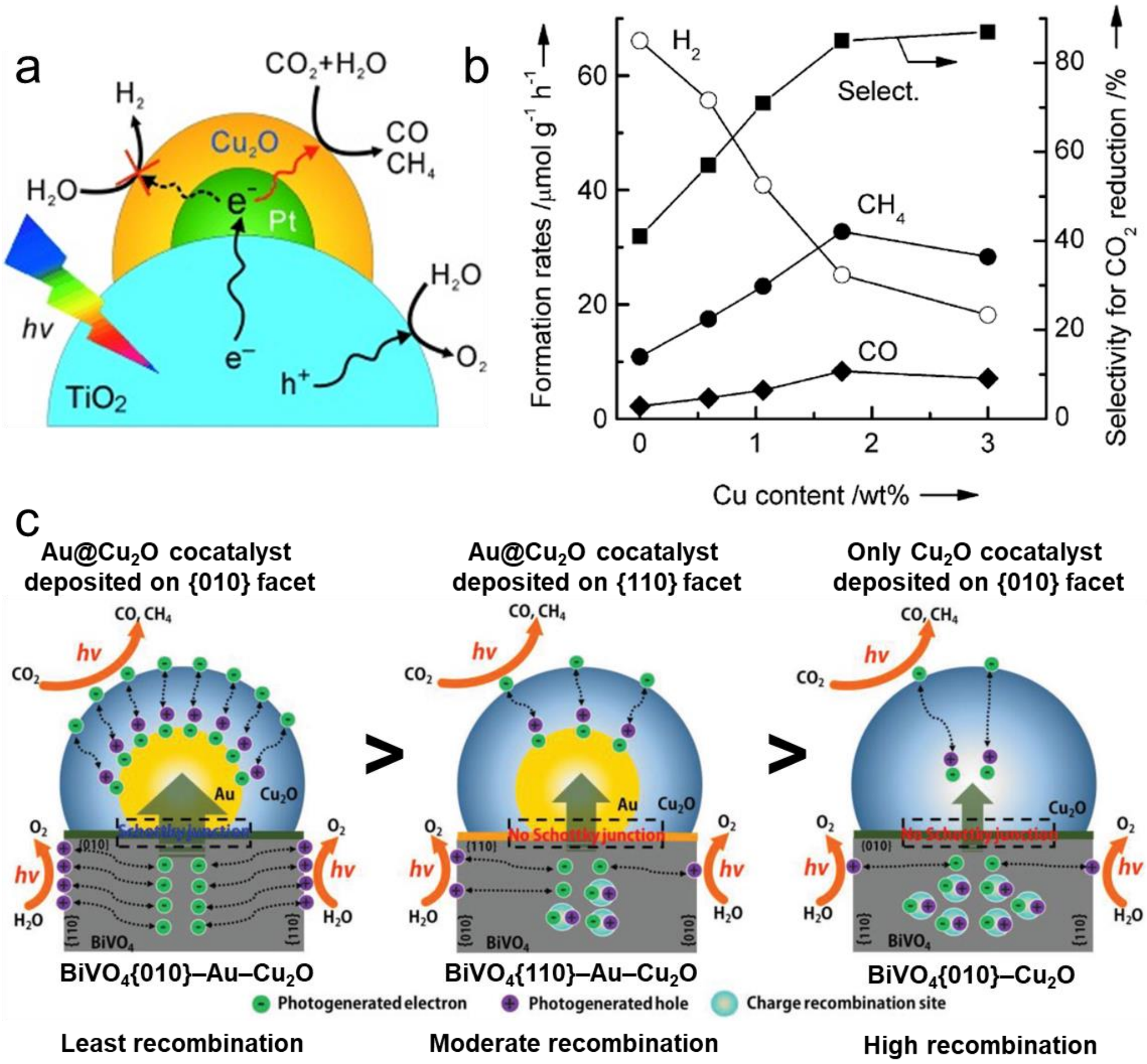

**Figure 27:** (a) Schematic mechanism for $Cu_2O/Pt/TiO_2$-*xh* hybrid system and (b) its product selectivity with respect to the Cu content in the hybrid catalysts for the photoreduction of $CO_2$ with $H_2O$. Reproduced with permission from ref. [208], Copyright 2013, Wiley-VCH. (c) Scheme to describe the carrier migration behavior from $BiVO_4$ to $Cu_2O$ with and without Au particles deposited on the different facets of $BiVO_4$. Reproduced with permission from ref. [210], Copyright 2018, Wiley-VCH.

An efficient Schottky junction was formed in the $BiVO_4$\{010\}–Au interface, facilitating the extraction of the photogenerated electrons from $BiVO_4$ and their successful transfer to $Cu_2O$, thus improving the charge separations of the system (Fig. 27c). $BiVO_4$\{010\}-Au-$Cu_2O$ showed a $CH_4$ production rate equal to 3.14 $\mu molg_{cat}^{-1}h^{-1}$ (1 $\mu molm^{-2}h^{-1}$ in terms of surface area), which was three and five times higher than $BiVO_4$\{110\}-Au-$Cu_2O$ and $BiVO_4$\{010\}-$Cu_2O$, respectively. The Schottky junction between $BiVO_4$ and Au not only facilitated a better charge separation but also prevented the photocorrosion of $Cu_2O$ by neutralizing the holes, which resulted in a high photostability of the entire system. Despite the good results, it is worth mentioning that such low $CH_4$ and CO production rates can be associated with surface carbon contamination. Isotope labeled $^{13}CO_2$ photoreduction was not reported either in this study, thus some doubts remain on the origin of the detected products. Recently, Lee et al. showed that different surface ligands on the $Cu_2O/TiO_2$ photocatalyst surface alters the binding strength of reaction intermediates resulted in different product selectivities during the gas-phase photocatalytic $CO_2$ reduction [217]. Therefore, authors should consider any overestimation of the activity for photoreduction of $CO_2$. In general, previous case studies provided an alternative strategy for developing hybrid photocatalytic systems, which can reduce the $H_2$ evolution reduction while improving the formation of solar fuels with suppressed photocorrosion.

## 5. Conclusion and remarks

Substantial progress in the understanding of synthetic procedures and growth mechanisms of the faceted $Cu_2O$ crystals opens the door to studying their facet-dependent properties. New physicochemical properties have been discovered by studying the facet-dependent properties of $Cu_2O$ crystals, which are different from traditional viewpoints. In particular, facet-engineering of $Cu_2O$ nanocrystals provides the opportunity to tune their surface atomic arrangements and light absorbing properties and induce an increased separation of charge carriers on different anisotropic facets, which will result in an enhanced photocatalytic activity.

Comparison of surface energies of $Cu_2O$ crystallographic facets reveals that the order is as follows: $\gamma\{1\ 0\ 0\} < \gamma\{1\ 1\ 1\} < \gamma\{1\ 1\ 0\} < \gamma\{h\ k\ l\}$. Preparation of 1D, 2D, and 3D $Cu_2O$ nanostructures have been reviewed while highlighting the role of a surfactant, solution pH, and different inorganic ions in various synthetic chemical methods. Although obtaining such well-defined morphologies requires the use of templating and capping agents, their carful removal is crucial for guaranteeing the subsequent photocatalytic activity. Clean surfaces must be obtained, and they proved to not only maximize the activity but also ensure the absence of contaminants which can be released during the reaction and may alter the product distribution or can be involved as sacrificial electron donors leading to artefacts. Afterward, suitable steps have been suggested for efficiently removing the surfactants without distorting the facet and chemical compositions of the $Cu_2O$ nanocrystals.

This precise control of the facet enable unique properties such as facet-selective adsorption of capping agents and facet-selective photodeposition of oxidation and reduction cocatalyst. It was also identified that pure $Cu_2O$ nanocrystals showed facet-dependent optical properties. Furthermore, when core-shell nanocrystals were formed with a plasmonic core and $Cu_2O$ shell, a large red-shift for the SPR peak of the plasmonic core was observed, which was due to the differently exposed facets of the shell and their corresponding refractive indices or dielectric constants. Such interesting tuning of the optical properties of hybrid core-shell nanocrystals can provide a highly efficient photocatalysts, absorbing from the UV-Vis to NIR region of the solar spectrum.

Discussion on the necessary figures of merit in photocatalysis were also reviewed in order to avoid any mistakes during the demonstration of the photocatalytic performances. We discussed that the enhanced photocatalytic activity is directly related to: 1) efficient adsorption and activation of reactants on the catalyst surface, 2) upon exposure of light, well separation of photogenerated electron–hole pairs and their smooth migration to the active catalytic sites via the exposed crystal facet to carry out the photo-redox reaction. Then special emphasis was put on well-defined hybrid $Cu_2O$-based nanostructures, which showed enhanced photocatalytic performance for dye degradation, organic reactions, $CO_2$ reduction, and $H_2$ production, compared to the pristine $Cu_2O$ nanostructures. The possible charge transfer mechanisms and the product selectivity were also illustrated for different cases. We particularly emphasized the potential strategies for efficient consumption of photogenerated holes for oxidation reactions, thus reducing the self-photooxidation of $Cu_2O$ counterparts, which, in turn, increased the photostability. We hope that the presented in-depth discussion and suggestions will contribute to the discovery of new synthetic strategies to enable the preparation of diverse hybrid $Cu_2O$ nanostructures and the fundamental investigation of the facet-dependent properties on nanoscale. This may finally enable the design of low cost and more efficient $Cu_2O$-nanostructures for solar to fuel conversion technologies.

**Declaration of Competing Interest**

The authors declare that they have no known competing financial interests or personal relationships that could have appeared to influence the work reported in this paper.

**Acknowledgements**

M.B. and P.F. acknowledge financial support from European Community (projects H2020 - RIA-CE-NMBP-25 Program - Grant No. 862030 - and H2020-LC-SC3-2019-NZE-RES-CC - Grant n 884444), INSTM consortium and ICCOM-CNR. S. R and A.N. gratefully acknowledge the support of the Czech Science Foundation (GACR) through the project no. 20-17636S, the Ministry of Education, Youth and Sports of the Czech Republic through the projects ERC CZ no.